\documentclass[12pt,preprint]{aastex}
\received{}
\revised{}
\accepted{}

\shorttitle{Simulating the X-ray Forest}
\shortauthors{Fang, Bryan, \& Canizares}

\begin{document}

\title{Simulating the X-ray Forest}

\author{Taotao Fang, Greg L. Bryan\altaffilmark{1} AND Claude R.Canizares}
\affil{Department of Physics and Center for Space Research}
\affil{Massachusetts Institute of Technology}
\affil{NE80-6081, 77 Massachusetts Avenue, Cambridge, MA 02139}
\altaffiltext{1}{Hubble Fellow}
\email{fangt@space.mit.edu, gbryan@alcrux.mit.edu, crc@space.mit.edu}

\begin{abstract}

Numerical simulations predict that a large number of baryons reside in intergalactic space at temperatures between $10^{5}-10^{7}$ K. Highly-ionized metals, such as \ion{O}{7} and \ion{O}{8}, are good tracers of this ``{\it warm-hot intergalactic medium}'', or WHIM. For collisionally-ionized gas, the ionization fraction of each ion peaks  at some particular temperature (``peak temperatures''), so different ions can therefore trace the IGM at different temperatures. We performed a hydrodynamic simulation to study the metal distributions in the IGM. We then draw random lines-of-sight across the simulated region and synthesize resonance absorption line spectra in a similar way to simulating the Ly$\alpha$ forest. By studying the distribution functions of H- and He-like O, Si and Fe in a collisionally-ionized IGM and comparing with semi-analytic results based on the Press-Schechter formalism, we find: (1) ions with higher peak temperatures (for instance, \ion{Fe}{26}) tend to concentrate around virialized halos, which can be well described by the Press-Schechter distribution, ions with lower peak temperatures are found both in small halos (such as groups of galaxies) and in filaments; (2) lower peak temperature ions are more abundant and should be easily observed; (3) peculiar velocities contribute a significant part to the broadening of the resonant absorption lines.

\end{abstract}


\clearpage

\section{Introduction}

Cosmological N-body simulation plays an important role in studying cosmic structure formation and evolution. In this methodology, initial conditions are first set at some early epoch based on the history of the early universe and the linear growth theory of initial perturbations. N-body simulations are then used when cosmic structure grows into the non-linear regime (for a review see \citealp{ber98}). The earliest cosmological N-body simulations involved only a simple system with a dissipationless dark matter component, and the evolution of such a system depends solely on gravity (see, e.g., \citealp{pee70,gta79}). An important step was to incorporate the gas dynamics for baryonic matter (see, e.g., \citealp{cos94,koc94}. The evolution of this baryonic matter depends not only on gravity but also on heating and cooling processes. This step turned out to be quite successful; for example it reproduced the observed statistical properties of the Ly$\alpha$ forest in spectra of high redshift quasars (see, e.g., \citealp{cmo94,zan95,zan97,mbm00}). In addition, numerical simulations based on the CDM models successfully fit the Press-Schechter distribution of galaxy clusters. Combined with observations (see, e.g., \citealp{har91} for {\sl Einstein} data, \citealp{bcs97} for {\sl ROSAT} data and \citet{hen97} for {\sl ASCA} data), this method can be used to constrain cosmological parameters, such as $\Omega$ and $\sigma_{8}$ (see, e.g., \citealp{ecf96,fbc97}).  

It is of particular interest to study the evolution of baryonic matter because it addresses an important question: why is there an apparent decrease in baryon density from high to low redshifts (see, e.g., \citealp{fhp98})? Numerical simulations indicated that the absence of baryons at low redshifts represents a yet undetected IGM: the ``{\it warm-hot intergalactic medium}'', or WHIM \citep{dco00}. For example, \citet{cos99} and \citet{dco00} found that the average temperature of baryons is an increasing function of time, with most of the baryons at the present time having a temperature in the range of $10^{5}-10^{7}$ K. Although invisible in the optical band, the highly ionized metals in WHIM gas might be detectable in UV and X-ray bands. Recent discovery of \ion{O}{6} absorption lines in the UV spectra of QSOs (see, e.g., \citealp{tls97,tsj00}) provided strong support for this, and implied a ``{\it significant baryon reservoir}'' at low redshifts. 

Since highly-ionized metals, such as \ion{O}{7} and \ion{O}{8}, are also good tracers of WHIM gas at $10^{5}-10^{7}$ K, in this paper we study the metal distributions in the IGM through hydrodynamic simulation. The idea is similar to that discussed in \citet{plo98}, \citet{hgm98} and \citet{fca00}: highly ionized metals would introduce absorption features in the X-ray spectrum of a distant quasar --- the ``{\it X-ray Forest}''. This paper has the following goals: (1) compare predictions of the X-ray forest from semi-analytic models to those from numerical simulations, this is important because each technique has its own strengths and weakness and by contrasting the two, we can determine which results are robust; (2) compute predictions for a wider range of ions, including \ion{O}{7}, \ion{O}{8}, \ion{Si}{13}, \ion{Si}{14}, \ion{Fe}{25}, \ion{Fe}{26}; (3) explore the nature of the gas that gives rise to the absorption, including its thermal state and its location in clusters, groups or filaments. The method we use is in analogy to the method in simulating the Ly$\alpha$ forest (see \citealp{zan97} for a complete discussion on formalism), and is briefly outlined here: first, random lines-of-sight (LOS) are drawn across the simulated region, and temperature, density and velocity are obtained at each point along every LOS to calculate the optical depth due to the absorbing ions; every possible line feature is then examined with a certain opacity threshold and for each identified line, we fit a Gaussian profile to obtain its equivalent width ($W$), the Doppler $b$-parameter and column density ($N^{i}$). Here $N^{i} = N(X^{i})$ is the column density of ion $X^{i}$. We then obtain the X-ray forest distribution function (XFDF) by averaging a total of $5,000$ LOS and compare it with those obtained by the semi-analytic calculations in \citet{fca00}. Similar work was also done before by \citet{hgm98}, based on a different ionization mechanism.

This paper is arranged as follows: section~\ref{sec:ns} is a brief introduction to the numerical simulation, in section~\ref{sec:sa} we generate a simulated spectrum from the numerical simulation data and fit each individual line. We get the statistical properties of the line distribution and compare them with the Press-Schechter formalism in section~\ref{sec:sp}. Section~\ref{sec:sd} is a summary and discussion.

\section{Numerical Simulation \label{sec:ns}}

\subsection{Method \label{sec:sm}}
We numerically simulated a cubic region of $100 h^{-1}$ Mpc, or 150 Mpc with $h=0.67$ on one side at $z ~\sim 0$. We model the dark matter with particles (as in a standard N-body code) but use the technique of {\it adaptive mesh refinement} (AMR) to follow the baryons. This method is designed to provide adaptive resolution combined with a modern, shock-capturing Eulerian hydrodynamics scheme. In this technique, a uniform mesh (``{\it parent}'' grid) covers the entire computational volume, and in regions that require higher resolution, a finer subgrid (``{\it child}'' grid) is added. By repeating this process recursively, high resolution can be achieved in small, high-density regions. To communicate between {\it parent} and  {\it child} grids, boundary conditions go from coarse to fine, and the improved solution on the {\it child} grids is used to update the coarse {\it parent} grid. The advancing of the grid hierarchy is done automatically and dynamically, so interesting features can be followed at high resolution without interruption.  We refer to \citet{bry96} and \citet{nbr99} for further details.

In this simulation the dark matter particle mass was $5 \times 10^{11}\ M_{\odot}$ and the smallest cell size (in the high density regions) was $49h^{-1}$ kpc. To generate spectra, we created a uniform grid which contained $512 \times 512 \times 512$ cells, with each cell having a size of $\sim 195h^{-1}$ Kpc, or $19.5\ km\ s^{-1}$ in velocity space. The simulation itself is a flat low-density cold dark matter model (LCDM) with $\Omega_{0} = 0.3$, $\Omega_{\Lambda} = 0.7$ and $\Omega_{b} = 0.04$. The amplitude of the initial power spectrum was fixed so that the {\it rms} overdensity in a spherical tophat with radius $8h^{-1}$ Mpc, $\sigma_8$, is $0.9$. We use a scale-invariant power spectrum with a primordial index of $n=1$. The initial cosmological redshift is $z=30$.

\subsection{Gas Distribution \label{sec:gdis}}

\begin{figure}
\plotone{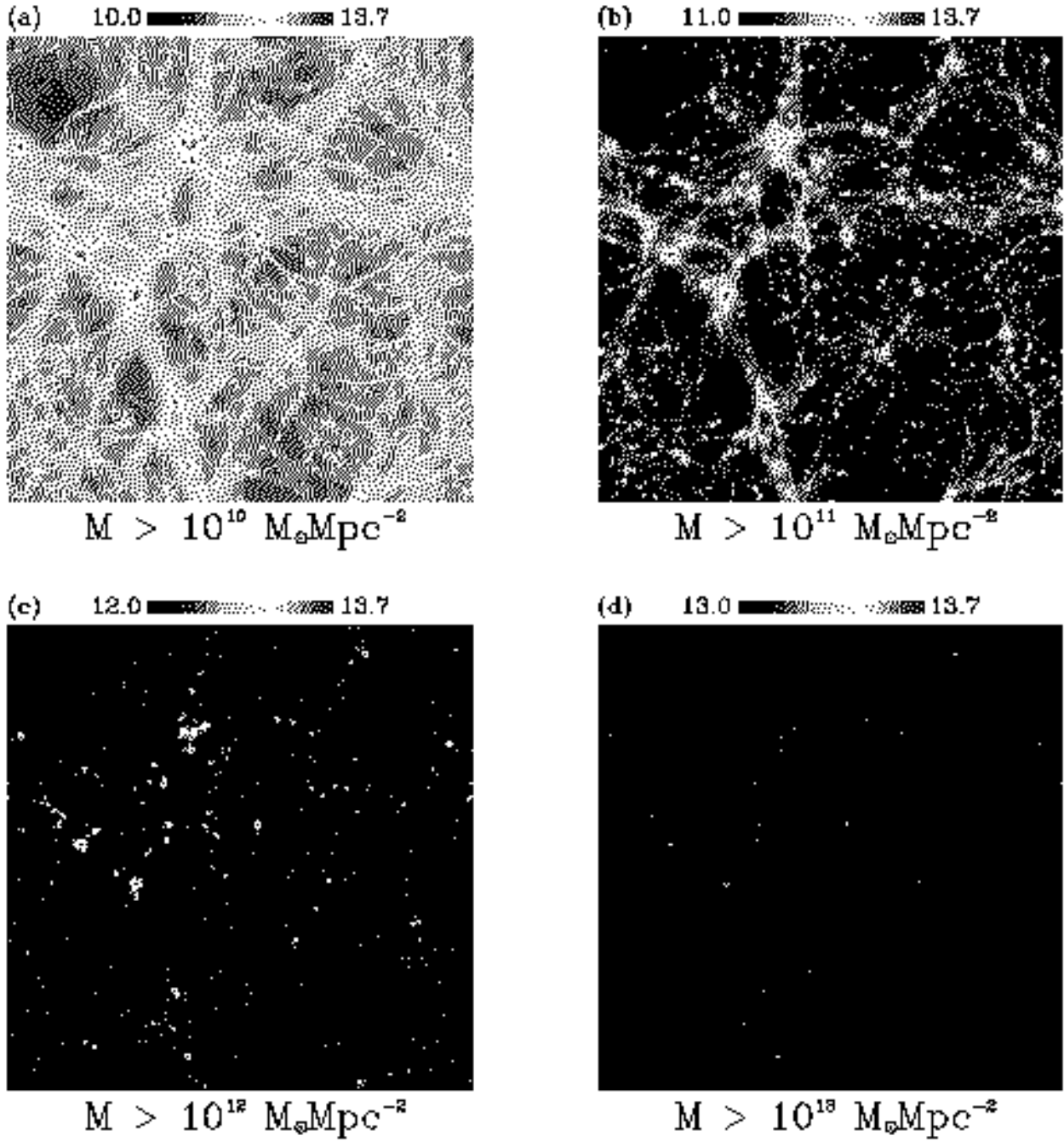}
\caption{The projected surface density ($M_{proj}$, in unit of $M_{\odot}Mpc^{-2}$) of the entire simulated region. Four panels are displayed with projected masses of $M_{proj}>10^{10}$, $10^{11}$, $10^{12}$ and $10^{13}\ M_{\odot}Mpc^{-2}$, respectively. The color bar on top of each plot indicates the color scale in logarithm. \label{f1}}
\end{figure}

Figure~\ref{f1} displays the projected baryonic density ($M_{proj}$, in unit of $M_{\odot}Mpc^{-2}$) of the entire simulated region. Four panels are displayed with projected masses of $M_{proj}>10^{10}$, $10^{11}$, $10^{12}$ and $10^{13}\ M_{\odot}Mpc^{-2}$, respectively. We can immediately distinguish four apparent features in Figure~\ref{f1}a: the red spots ($M_{proj} > 10^{13}\ M_{\odot}Mpc^{-2}$), which indicate regions with the highest densities; the yellow extended structures ($M_{proj} \sim 10^{12}-10^{13}\ M_{\odot}Mpc^{-2}$) surrounding the red spots; the green filamentary structures ($M_{proj} \sim 10^{10}-10^{12}\ M_{\odot}Mpc^{-2}$) connecting the high density regions and the dark region ($M_{proj} < 10^{10}\ M_{\odot}Mpc^{-2}$, for instance, on the top left corner) indicating regions with the lowest densities. By looking from panel (a) through (b) to (c) and (d) we find that as the cut-off $M_{proj}$ increases, most of filaments begin to disappear and high density regions start to concentrate only on several isolated red spots (for example, see panel c and d). These highly mass-concentrated regions indicate that collapsed, virialized objects, such as groups and clusters of galaxies, have formed there. Since overlap of these massive systems is rare along the projected direction, a projected surface density of $10^{12}$ or $10^{13}\ M_{\odot}Mpc^{-2}$ roughly corresponds to a group or cluster of galaxies with $M \sim 10^{12}$ or $10^{13}\ M_{\odot}$, respectively.

\begin{figure}
\plotone{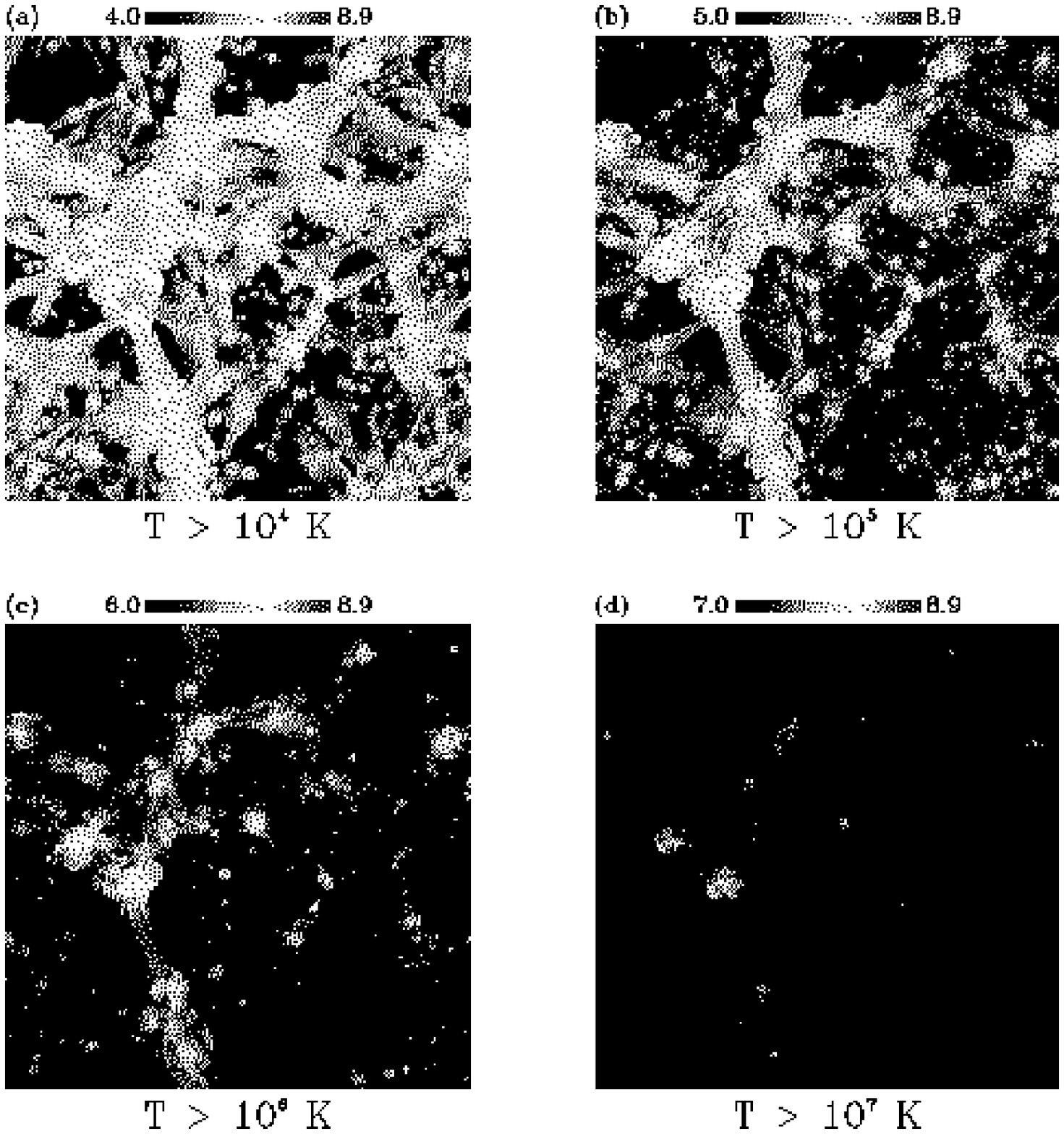}
\caption{The emission-weighted temperature distribution of the simulated region. The four plots display the temperature distributions in four ranges: $T>10^{4}$, $10^{5}$, $10^{6}$ and $10^{7}$ K. The highest temperature in our simulation is $8.09 \times 10^{8}$ K. The temperatures below about $2\times10^{4}$ K are not real. \label{f2}}
\end{figure}

Figure~\ref{f2} shows the emission-weighted temperature distribution of the simulated region. The four plots display the temperature distributions in four ranges: $T>10^{4}$, $10^{5}$, $10^{6}$ and $10^{7}$ K. The temperatures below about $2\times10^{4}$ K are not real because the simulation does not include an ionizing radition field which would heat the gas up. The highest temperature in our simulation is $8.09 \times 10^{8}$ K. As shown by figure 2a, 2b, 2c, most of the simulated regions are filaments with $10^{4}$ K $< T < 10^{6}$ K. The intersections between filaments show higher temperatures with $T > 10^{6}$ K. Comparing Figure~\ref{f2} with Figure~\ref{f1}, we find that high temperature regions correspond to high density regions. For instance, comparing panel c in both figures, a projected surface density of $10^{12}\ M_{\odot}Mpc^{-2}$ implies a group of galaxies with $M \sim 10^{12}\ M_{\odot}$ and $T \sim 10^{6}$ K, as indicated in Figure~\ref{f2}c.

\placefigure{f1}
\placefigure{f2}

To ease the load of numerical computing, no atomic and radiative processes are taken into account here, so high gas temperature is mainly due to adiabatic compression and shock heating. Gas heating and cooling processes will play an important role in the cold gas (and stars) in galaxies, but should play only a minor role for hot gas. \citet{dco00} compared an AMR simulation including only adiabatic process (model B1 in their paper) with an AMR simulation including several gas heating and cooling processes (model B2). These processes include H, He and metal cooling, photoionization and star formation \& feedback. They checked the baryon fraction of WHIM gas and found that both numerical simulations gave consistent results at $z=0$ ($\Omega_{WHIM}/\Omega_{b} = 0.32$ for model B1, and $\approx 0.3$ for model B2). By comparing the two AMR simulations with several other numerical simulations they found that gravitational shock heating of gas falling onto large scale structures is the dominant heating mechanism for WHIM gas. Most of the higher temperature IGM ($T > 10^{7}$ K) is associated with hot intracluster gas (see Figure~\ref{f1} and ~\ref{f2}). The major cooling mechanisms for intracluster gas are thermal bremsstrahlung radiation and X-ray line emission. The cooling time scale is approximately given by \citep{sar88}
\begin{equation}
t_{cool} = 8.5 \times 10^{10} \left(\frac{n_{p}}{10^{-3}\ cm^{-3}}\right)^{-1}\left(\frac{T}{10^{8}\ {\rm K}}\right)^{\frac{1}{2}} yrs,
\end{equation} which is longer than the Hubble time. Here $n_{p}$ is the proton density and typically in the order of $10^{-3}\ cm^{-3}$. Based on the above considerations, we find that excluding gas heating and cooling in our simulation is a reasonable approximation for an IGM with $T>10^{5}$ K.

\subsection{Metal Distribution \label{sec:md}}

Little is known about the metallicity of the IGM. The metal abundance could be as low as $10^{-3}Z_{\odot}$ in the low column density Ly$\alpha$ forest systems \citep{lss98} or as high as $1Z_{\odot}$ in the interstellar medium (ISM) of some early-type, X-ray luminous galaxies \citep{mom00}. Since we will compare numerical simulation with the semi-analytic model in our previous paper \citep{fca00}, we assume uniform metalicity and use the metallicities observed from the intracluster medium, e.g., $0.5Z_{\odot}$ for oxygen, silicon and $0.3Z_{\odot}$ for iron \citep{mla96}. Table~\ref{t1} lists parameters for the ions and lines considered here.

\placetable{t1}

\begin{figure}
\plotone{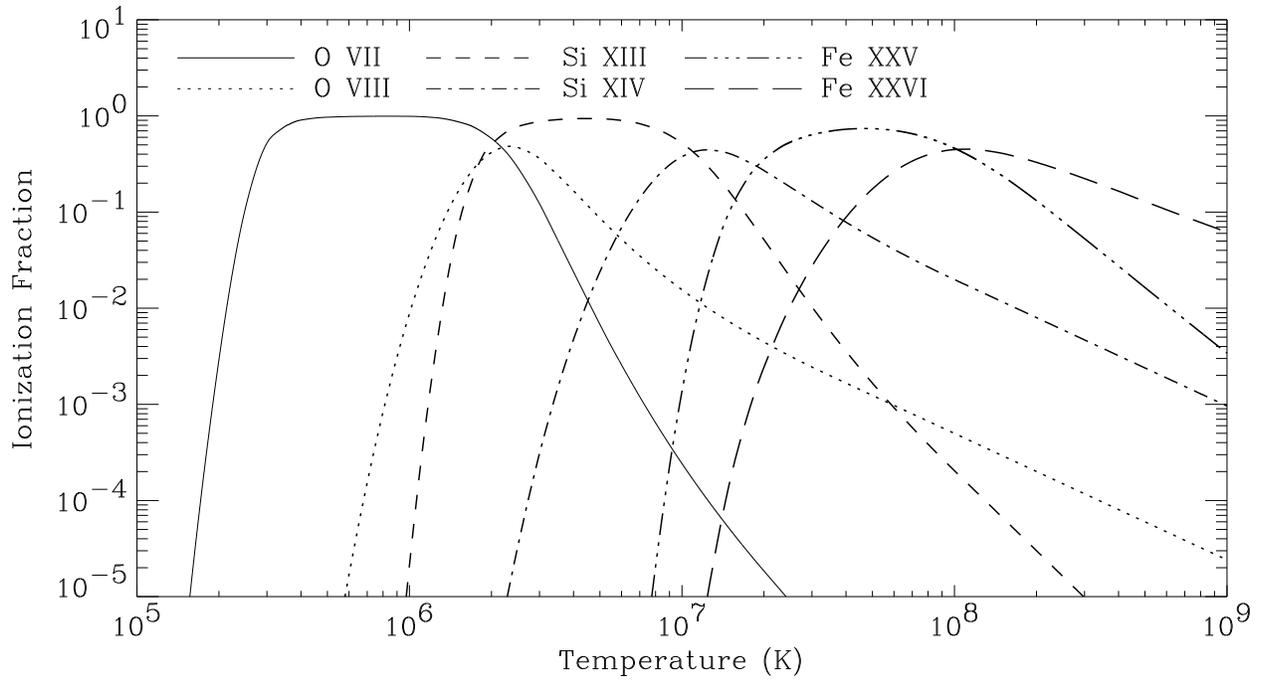}
\caption{Ionization fractions from \ion{O}{7} (solid curve) to \ion{Fe}{26} (long-dashed curve). \label{f3}}
\end{figure}

To investigate X-ray absorption in the spectrum of a background source, we need to understand the distribution of ionization states of metals in the IGM. To obtain the ion density, we assume the gas is in collisional equilibrium and we adopt the ionization fractions from \citet{mmc98}. We select helium- and hydrogen-like oxygen, silicon and iron. The peak temperatures of these ion species range from  $10^{5}$ K to $10^{9}$ K (see Figure~\ref{f3}), which probe different temperature regions of the IGM. 

\placefigure{f3}

Figure~\ref{f4} through Figure~\ref{f9} display the projected column density, $N^{i}$, distributions of all the six ions. In each figure, the ion column density distributions are plotted in four different panels with $N^{i} > 10^{12}$, $10^{13}$, $10^{14}$ and $10^{15}\ cm^{-2}$, respectively. The color bar on top of each panel indicates the logarithm scale from the minimum to the maximum column densities in each panel. 

\placefigure{f4}
\placefigure{f5}
\placefigure{f6}
\placefigure{f7}
\placefigure{f8}
\placefigure{f9}

By examining the six figures, we find that the distributions of these ions are closely related to the density (Figure~\ref{f1}) and temperature(Figure~\ref{f2}) distributions. In general, column density distributions of metals follow the projected density distribution in Figure~\ref{f1}: regions with higher baryon densities also have higher metal densities. Due to the fact that different ions have different peak temperatures, there are differences among the distributions of these ion species. For instance, Figure~\ref{f2}d shows that many of the filaments that connect high density regions are shock-heated to temperatures above $10^{5}$ K. Since \ion{O}{7} has a peak temperature between $3\times10^{5}$ K and $2\times10^{6}$ (Figure~\ref{f3}), we find that a large amount of \ion{O}{7} exists in filaments with a column density even higher than $10^{15}\ cm^{-2}$ (Figure~\ref{f4}c). On the other hand, \ion{Fe}{26} has a peak temperature of over $10^{8}$ K. In Figure~\ref{f9} we find that almost no \ion{Fe}{26} exists in filaments, and it all concentrates around those collapsed intersections that appear in Figure~\ref{f1}. By comparing these two extremes, we find that generally \ion{Fe}{26} follows the distribution of virialized objects while most of \ion{O}{7} exists in filaments, and the distributions of other ion species vary between those of \ion{Fe}{26} and \ion{O}{7}.

\section{Spectral Analysis \label{sec:sa}}

The methodology we adopt is straightforward: a random line-of-sight (LOS) is drawn across the simulated region; temperature, baryon density and velocity are obtained at each point along the LOS; the ion density is calculated based on the metallicity and ionization fraction; the synthesized spectrum is then obtained from the optical depth along the LOS; finally, by fitting the resonant absorption lines in the ``{\it simulated}'' spectrum we obtain ion column densities ($N^{i}$) and the Doppler parameters ($b$). This method is similar to the method used in the Ly$\alpha$ forest; we refer to \citet{zan97} for a complete discussion and formalism.   

\subsection{Gas Distribution along the LOS \label{sec:gdal}}

To investigate the gas distribution along the LOS, we draw random lines across the simulated region. Each LOS has 512 points. The separation between each point is $0.195 h^{-1}$ Mpc, or $\Delta z = 6.5 \times 10^{-5}$ in redshift space. This roughly corresponds to an energy resolution of $0.04\ eV$ around $0.6\ keV$ (oxygen line region), or $0.44\ eV$ around $6.7\ keV$ (iron line region). We obtain temperature, peculiar velocity (coherent or turbulent) and gas density at each point. For demonstration, we show their distributions along one LOS in Figure~\ref{f10}: Figure~\ref{f10}a shows the temperature distribution, Figure~\ref{f10}b gives the velocity distribution and Figure~\ref{f10}c gives the baryon number density distribution. 

\placefigure{f10}

The dotted line in Figure~\ref{f10}c is the mean baryon density of the universe 
\begin{equation}
\langle\rho\rangle = 1.124 \times 10^{-5} \Omega_{b} h^{2}\ cm^{-3} = 2.136 \times 10^{-7} \ cm^{-3}
\end{equation} Here $\Omega_{b}$ is the baryon density and we adopt a value of $\Omega_{b} h^{2} = 0.019$ from the standard theory of big bang nucleosynthesis (BBN) (see, e.g., \citealp{bnt00}). Most of the regions along this random LOS are underdense regions, with several peaks indicating overdense regions. The largest overdense region occurs at a distance of approximately $50\ Mpc$, with $\delta \rho = \rho/\langle\rho\rangle \approx 20$. Given the temperature $T \sim 10^{6}-10^{7}$ K (see the corresponding position in Figure~\ref{f10}a), this region might correspond to a small cluster or group of galaxies.

We notice that one region (at $\sim 110$ Mpc) shows both a very high temperature ($2-5\times 10^{7}$ K) and a high peculiar velocity, while this region is still underdense. This indicates that the LOS passed near a collapsed structure. The shock waves from collapsed structures can propagate deeply into low density regions, which cause a strong divergence of the peculiar velocity, and in turn a high temperature in this underdense region.

The density peaks of different ion species are closely related to peaks of their ionization fraction curves (see Figure~\ref{f11}). For instance, the peak temperature of \ion{Si}{14}, \ion{Fe}{25} and \ion{Fe}{26} are all above $10^{7}$ K, and since the only region with such a high temperature along this LOS is at $\sim 110$ Mpc (see Figure~\ref{f10}a), these ion species have just one density peak around that region (the bottom three panels of Figure~\ref{f11}), although this region is below the mean density of the universe.

\placefigure{f11}

\subsection{Simulated Spectra \label{sec:sg}}

Considering the X-ray spectrum of a background quasar located at redshift $z_{0}$, the frequency of a photon emitted at $\nu_{0}$ will be $\nu = \nu_{0} (1+z_{0})^{-1}$ when the photon reaches the observer. The optical depth at frequency $\nu$ (observer frame) is \citep{spi78}
\begin{equation}
\tau(\nu) = \int_{0}^{z_{0}} n_{i}\left(z\right)\sigma_{i}\left(\nu^{\prime}\right)\frac{dl}{dz}dz
\end{equation} where $n_{i}(z)$ is the comoving number density of absorbing ion $i$ at a redshift of $z$ and $\sigma_{i}$ is the ion absorption cross section at frequency $\nu^{\prime}$. Besides the Hubble velocity, the peculiar velocity (both coherent and turbulent) also contributes to the line broadening and produces a shift at the line center. Assuming a peculiar velocity of $v_{pec}$ along the line-of-sight (LOS), the frequency at redshift $z$ is
\begin{equation}
\nu^{\prime} = \nu \left(1+z\right) \left(1+\frac{v_{pec}}{c}\right)
\end{equation} To the first order $(v_{pec}/c)$ approximation the optical depth at frequency $\nu$ (observer frame) is (\citet{zan97})
\begin{equation}
\label{eq:tau}
      \tau(\nu) =
      \frac{\sigma_{0}}{\sqrt{\pi}\nu_{0}}\frac{c}{b_{T}}\int_{0}^{z_{0}}
      n_{i}\left(z\right)exp\left\{-\left[(1+z)\frac{\nu}{\nu_{0}}-1+\frac{v_{pec}}{c}\right]^{2}\frac{c^{2}}{b_{T}^{2}}\right\}\frac{dl}{dz}dz 
\end{equation} Here $\sigma_{0} = (\pi e^2/m_{e}c)f$ is the absorption cross section at the line center ($f$ is the upward oscillator strength) and $b_{T}$ is the velocity dispersion due to the thermal motion.

By integrating over a distance of $150 Mpc$ ($z_{0} \approx 0.034$ in our cosmological model) we synthesize the absorption spectrum. The solid lines in Figure~\ref{f12} show the transmission, $F = \exp\left(-\tau\right)$, from \ion{O}{7} to \ion{Fe}{26} (top to bottom), where $\tau$ is calculated from equation (\ref{eq:tau}). The translation from redshift ($z$) to wavelength ($\lambda$) is simply 
\begin{equation}
z = \frac{\lambda}{\lambda_{i}} - 1
\end{equation} where $\lambda_{i}$ is the ion's rest wavelength.

\placefigure{f12}

Comparing Figure~\ref{f11} and Figure~\ref{f12} we can see that absorption features are created by density peaks. Along this random LOS we find that most of the lines are very weak ($\tau < 0.01$), except for some \ion{O}{7} lines, where the largest optical depth at the line center is almost $\tau \approx 2$. The densities of \ion{Fe}{26} are so small (less than $10^{-14}\ cm^{-3}$) that no absorption feature is displayed in the bottom panel of Figure~\ref{f12}.

\subsection{Line Fitting \label{sec:lf}}

To exactly measure absorption lines we need to fit the synthesized spectrum with a Voigt line profile. Numerically such a procedure is difficult to achieve. However, the line profile can be well represented by a Gaussian line profile in our situation (see \citealp{zan97} for discussion). To simplify we adopt the ``{\it threshold-based}'' method of spectral analysis developed by \citet{mco96} and \citet{hkw96}. In this method, we follow the flux of a simulated spectrum. We first define a threshold $F_{t}$. An absorption feature is identified whenever the flux drops below ($1-F_{t}$) at wavelength $\lambda_{1}$ and rises above $1-F_{t}$ at wavelength $\lambda_{2}$ where $\lambda_{2} > \lambda_{1}$. In case of line blending between $\lambda_{1}$ and $\lambda_{2}$, different lines are identified whenever the slope of transmission curve changes sign and fitting is performed for each identified line. Here we see that $F_{t}$ actually represents the minimum central optical depth of lines which can be resolved by this method. The line equivalent width ($W$) is calculated by integrating ($1-F_{t}$) from $\lambda_{1}$ to $\lambda_{2}$. Taking into account redshift, the equivalent width is given by
\begin{equation}
        W =
        \int_{\lambda_{1}}^{\lambda_{2}}(1-e^{-\tau})\frac{d\lambda}{1+z} 
\end{equation}

The Doppler $b$-parameter is defined as the $b$-parameter of a Voigt profile which gives the same equivalent width with internal width $\Delta\lambda = \lambda_{2} - \lambda_{1}$ between $\lambda_{1}$ to $\lambda_{2}$. Following \citet{spi78} we have
\begin{equation}
      b = \frac{Wc}{2\lambda_{0}F(\tau_{0})}
\end{equation} Here $\tau_{0}$ is the optical depth at the line center, and
\begin{equation}
      F(\tau_{0}) =
      \int_{0}^{\infty}\left[1-exp\left(-\tau_{0}e^{-x^{2}}\right)\right]dx
\end{equation} is the curve-of-growth function. 

The ion column density is given by
\begin{equation}
\label{eq:Nb}
      N^{i} = \frac{\sqrt{\pi}\tau_{0}b}{\sigma_{0}\lambda_{0}}
\end{equation} Here $\sigma_{0}$ is the absorption cross section at the line center. 

We need to understand the effect of threshold on line identification before we analyze spectral properties. Equation~(\ref{eq:Nb}) shows that for a given threshold, there exists a minimum resolvable column density: the ``{\it threshold-based}'' method cannot resolve a line with flux above this threshold. The lower the threshold is, the lower the minimum column density is. To understand this effect, we fit the synthesized \ion{O}{7} and \ion{O}{8} spectra for the 5000 random line-of-sight (LOS), with different thresholds ranging from $F_{t}=10^{-5}$ to $F_{t}=0.8$. In Table~\ref{t2}, we list thresholds (the first column) and the minimum resolved column densities for \ion{O}{7} (the second column) and \ion{O}{8} (the third column). As we discussed in \citet{fca00}, the minimum detectable column density for oxygen, even with the future mission {\sl Constellation-X}, is $\sim 10^{15}\ cm^{-2}$, which means that a high threshold of $F_{t}=0.5-0.8$ is more realistic from an observational point of view. However, in order to understand the low density regions with low metal column densities and make our sample theoretically complete, in the following discussion we select the lowest threshold, $F_{t} = 10^{-5}$, to study spectral properties statistically. This also helps in comparing the numerical simulation with the semi-analytic calculations we performed in \citet{fca00}.

\placetable{t2}

Figure~\ref{f12} shows six spectra from the sample LOS in section \S\ref{sec:sg} ranging from \ion{O}{7} to \ion{Fe}{26}. The solid lines are the spectra from the numerical simulation and the dashed lines are from the fitted data by the ``{\it threshold-based}'' method.  Here $F_{t} = 10^{-5}$ (the dotted line). We identify five \ion{O}{7}, four \ion{O}{8},  five \ion{Si}{13}, two \ion{Si}{14}, one \ion{Fe}{24} and zero \ion{Fe}{26} absorption lines. 

The line parameters  are listed in Table~\ref{t3}. The first column is the ion species; the second column is the absorption line numbers; the third column gives the line redshift; the fourth column is the column density; the fifth column is the equivalent width; the final column gives the Doppler $b$-parameter. We can see that all lines are identified, even including those which are blended. For example, in the transmission spectrum of \ion{Si}{13} in Figure~\ref{f12}, lines 1 and 2 are blended with each other. Since the sign of the curve changes from line 1 (plus) to line 2 (minus), two lines are identified (see Table~\ref{t3}) with a separation of $\Delta z = 3.3 \times 10^{-4}$. Table~\ref{t3} shows that the equivalent widths of most absorption lines are so weak ($W < 0.1 eV$) that they are practically undetectable. However, this table does show some \ion{O}{7} lines with $W > 0.1 eV$, which might be detectable by future missions, such as {\sl Constellation-X}. The \ion{Fe}{25} line shows a large $b$ value, indicating that this line is generated in a high temperature region.

\placetable{t3}

\section{Spectral Properties \label{sec:sp}}

In this section we discuss the statistical properties of the absorption lines identified in our numerical simulation. We draw a total of 5000 random LOS across the simulated region. For each LOS we generate a simulated spectrum and fit it by the ``{\it threshold-based}'' method. Table~\ref{t4} lists the maximum and minimum value of column density and the Doppler $b$-parameter for each ion for $F_{t} = 10^{-5}$. In the following sections we will discuss column density and the Doppler $b$-parameter distributions of these lines.

\placetable{t4}

\subsection{Column Density Distribution \label{sec:cdd}}
  
In analogy to the Lyman-$\alpha$ forest system, we define $\partial^{2}P/\partial N^{i}\partial z$ as the number of absorption systems along the line of sight with a column density between $N^{i}$ and $N^{i}+dN^{i}$ per unit redshift. This is the so-called ``X-ray forest'' distribution function (XFDF).  The calculation of XFDF is straightforward: for each ion, the column density is divided into several bins between $10^{12}$ and $10^{17}\ cm^{-2}$; absorption lines are counted in each column density interval and then divided by $\Delta z$. Here $\Delta z = 3.4\times10^{-2}$ is the size of the simulated region in redshift space. The lower limit of the column density is selected arbitrarily; although a lower limit of $10^{12}\ cm^{-2}$ is undetectable, we still include it to make our sample theoretically complete; the upper limit is selected by noticing that no ions shows a column density larger than $10^{17}\ cm^{-2}$ in Table~\ref{t4}.  

Figure~\ref{f13} displays the distribution functions from \ion{O}{7} to \ion{Fe}{26}. The solid lines show all the helium-like ion species, while the dashed lines show all the hydrogen-like ion species. The main feature seen in this figure is that all curves display a sharp cutoff somewhere between $10^{16}$ and $10^{17}\ cm^{-2}$. The reason for this apparent drop at high column density end is two fold: first, systems with high column density generally correspond to high temperature regions (for example, compare Figure~\ref{f1} with Figures~\ref{f4} through \ref{f9}), and at high temperature the ionization fraction drops quickly and so does the ion column density; second, as we emphasized in section \S\ref{sec:md}, high column density systems are closely related to virialized objects, such as groups and clusters of galaxies, and the Press-Schechter theory predicts that the spatial density of virialized objects decreases exponentially as their masses (and temperatures) increase. Based on this explanation, the maximum ion column densities ($N^{i}_{max}$ in Table~\ref{t4}) should come from the centers of virialized objects which have temperatures close to the peak temperature of that ion. For instance, a rough estimation shows that for a galaxy cluster at temperature $5\times10^{7}$ K, the central column density of \ion{Fe}{25} is $\sim 4.7\times10^{16}\ cm^{-2}$, which is about the same order as the value shown in Table~\ref{t4}.

\placefigure{f13}

Figure~\ref{f13} also shows that the overall amplitude of the XFDF is a decreasing function of the ions' peak temperature: ions with low peak temperatures display higher amplitudes which implies more absorption lines (see table~\ref{t4}). This is not only due to the differences of the metal abundances, but also due to the fact that most of the baryons have temperatures ranging from $10^{5}-10^{7}$ K and the baryon fraction decreases as the temperature increases \citep{dco00}. This implies that the low peak temperature ions such as \ion{O}{7} and \ion{O}{8} are more abundant and should be easily observed.

\subsection{Press-Schechter Distribution \label{sec:psd}}

Returning to Figure~\ref{f1} and \ref{f2} we find that in general mass and temperature distributions follow each other: the low-density regions, such as filaments, show temperatures between $10^{4}$ and $10^{6}$ K; the high-mass regions, such as intersections between filaments, show higher temperatures with $T > 10^{6}-10^{7}$ K. High temperature and high density indicate that collapsed, virialized objects, such as groups and clusters of galaxies, might be formed in these intersection regions. As we emphasized before, due to the different peak temperatures of the ion species, these ions actually probe the different temperature regions of the IGM. For instance, \ion{O}{7} and \ion{O}{8}, which have peak temperatures around $10^{6}$ K, should follow the distributions of those filamentary structures; on the other hand, \ion{Fe}{25} and \ion{Fe}{26}, which have peak temperatures over $10^{8}$ K, should follow the distributions of those collapsed, virialized objects. Figures~\ref{f4} through \ref{f9} confirm this point.

Based on the formalism in \citet{fca00}, we compare our numerical simulation with the analytic model. The Press-Schechter function \citep{psc74} predicts the spatial distribution of virialized objects. We would expect that the analytic model and the numerical simulation should give consistent results for the distribution of the high-peak-temperature ions, such as \ion{Fe}{25} and \ion{Fe}{26}, because these ions follow the distribution of virialized objects; on the other hand, we would not expect the low-peak-temperature ions, such as \ion{O}{7} and \ion{O}{8}, would have the same behavior because most of these ions are distributed in filaments.

Before we compare the ion column density distributions, we first need to compare the spatial number density distribution of the virialized objects identified in our numerical simulation with that predicted by the Press-Schechter function. To identify the virialized objects in the simulated regions, the spherical overdensity algorithm \citep{cla96} is applied. In this algorithm, peaks in the density distribution are identified as the center of virialized objects; around each point, a sphere grows until the mean density inside the sphere reaches some critical value, $\Delta_{c}$. The mass center of this sphere is computed and used to grow a new sphere. This procedure is iterated to convergence. If the center-to-center distance of two identified objects is less than $3/4$ of the sum of their virial radii, we define these two objects as overlapping and the less massive object is removed (see \citealp{bno98}).

Figure~\ref{f14} displays the comparison between the numerical simulation (the filled square) and the analytic model (the sold line) for the number density of virialized objects. The analytic calculation is based on the LCDM cosmological model we adopt in the numerical simulation. The agreement is generally good, although the low-mass objects identified by the numerical simulation are systematically fewer than predicted. This is due in part to the fact that those low-mass objects are poorly resolved. However, there is also evidence in many numerical simulations indicating that the disagreement seems to be a general shortcoming of the PS method (see \citealp{lco94,gsp98}). \citet{gbf98} found that at $\sigma_{8} = 0.7$ or higher, the PS approach overestimates the number of small halos with $M<10^{14}h^{-1}\ M_{\odot}$, which is consistent with our result. This deficit has been associated with the merger events not accounted for by the PS formalism (\citealp{cme97,mon97}). By studying the clustering of less massive halos \citet{sto99} revised the PS function with a bias factor and obtained a good fit to numerical simulations for small halos as well as large halos. In our case the agreement in the high-mass region ($M>10^{14}h^{-1}\ M_{\odot}$) is very good.  

\placefigure{f14}

The techniques used to calculate the XFDF have been described in \citet{fca00}, so we only briefly summarize them here. The mass of a virialized halo is
\begin{equation}
M_{vir} = \frac{4}{3} \pi r^{3}_{vir} \rho_{crit} \Delta_{c}
\end{equation} Here $\rho_{crit}$ is the critical density of the universe, $r_{vir}$ is defined as the virial radius of a spherical volume within which the mean density is $\Delta_{c}\rho_{crit}$, and we adopt a conventional value of $\Delta_{c} = 200$. The mass-temperature relationship from \citet{ecf96} is used to convert between temperature and mass. Finally, to obtain the ion column density distribution within the halo we use the $\beta$-model of intracluster gas density distribution \citep{sar88} and assume a gas fraction of $f_{gas} = 0.2$.

Since ionization fractions are very sensitive to temperature, we also investigate the assumption that virialized halos are isothermal. In Figure~\ref{f15} we plot the normalized radial temperature profile for several identified clusters in the simulation. We find the overall radial temperature profile can be well represented by (the solid line)
\begin{equation}
\frac{T}{T_{vir}} = 1.1 \exp\left(-0.8\frac{r}{r_{vir}}\right)
\end{equation} Here $T_{vir}$ is the virial temperature of clusters. The computed ion column density using the PS formalism takes account of this temperature variation along the integral path.

\placefigure{f15}

Figure~\ref{f16} shows the XFDF predicted from the Press-Schechter approach for all the six ions (solid line in each plot). The filled-dotted line displays the distributions from the numerical simulation. The most striking feature is that the predicted distributions fit the simulation very well at high column density regions, specifically at $N_{i} > 10^{16}\ cm^{-2}$ for both oxygen species and at  $N_{i} > 10^{15}\ cm^{-2}$ for silicon species and \ion{Fe}{25}. For \ion{Fe}{26}, both distributions agree very well in the whole column density range ($10^{12} - 10^{16}\ cm^{-2}$). This result can be well understood in the framework we describe at the beginning of this section: high-peak-temperature ions, such as \ion{Fe}{26}, follow the distribution of collapsed, virialized halos well described by the PS formalism. For low-peak-temperature ions, the high column densities also come from virialized structures, while those with low column density are distributed in those filaments which can not be described by the PS distribution. Thus except for \ion{Fe}{26}, at low column density the PS distributions are systematically lower than in the numerical simulation.

\placefigure{f16}

For high column density \ion{O}{7} and \ion{O}{8} ($N > 1-2\times 10^{16}\ cm^{-2}$) we find that the PS distributions are slightly higher than the distributions from the numerical simulation. This is because the high column density oxygen lies in virialized halos with a temperature around $10^{6}$ K, which corresponds to a mass between $10^{12}$ to $10^{13}M_{\odot}$. As we find in Figure~\ref{f14}, the PS formalism overestimates the number of small halos, so the XFDFs for \ion{O}{7} and \ion{O}{8} at $N > 10^{16} cm^{-2}$ are also overestimated.

\subsection{Doppler Parameter Distribution \label{sec:dpd}}

Doppler $b$-parameter governs the line width and can be affected by several factors, such as gas temperature (thermal motion), the Hubble flow and the  velocity structure (see equation [\ref{eq:tau}]). We follow the approach used in the Ly$\alpha$ forest studies (\citealp{bma99,bma00}) and distinguish two sources of line broadening: $b_{T}$, the thermal broadening; $b_{prof}$, a convolution of the density profile with the peculiar velocity. The total $b$-parameter is
\begin{equation}
b = \sqrt{b_{T}^{2} +  b_{prof}^{2}}
\end{equation} 

Figure~\ref{f17} shows the $b$-parameter distribution function for six ions (the filled circles). All six ions exhibit a low cutoff to the Doppler parameter. This cutoff is also listed in Table~\ref{t4}. The cutoffs are primarily due to the cutoff temperatures for each ionization fraction (see figure~\ref{f3}). For instance, the minimum temperature we used for \ion{O}{8} is $T_{cut}=6.31\times10^{5}$ K, so
\begin{equation}
b_{cut} \approx b_{T} = \sqrt{\frac{2kT_{cut}}{Am_{p}}} =  25.55\ \ km\ s^{-1}
\end{equation} which is about the same value listed in Table~\ref{t4}. However, this also implies the fact that the thermal motion may play a dominant role in line broadening ($b_{prof} \approx 0$ and $b \approx b_{T}$). 

\placefigure{f17}

To understand the role of different line broadening mechanisms, we fit the $b$-parameter distribution with a Gaussian (the dashed line in Figure~\ref{f17})
\begin{equation}
f \propto \exp \left[-(b-\bar{b})^2/2\sigma^{2}\right]
\end{equation} Here $\bar{b}$ is the mean velocity and $\sigma$ is the standard deviation of a Gaussian curve. Both parameters are listed in the column 2 \& 3 of Table~\ref{t5}. By comparing $\bar{b}$ for different ions we can find that 
\begin{equation}
\bar{b} \propto \left(\frac{T_{peak}}{A}\right)^{\frac{1}{2}}
\end{equation} Here $T_{peak}$ is the peak temperature for different ions shown in Figure~\ref{f3}. This confirms that most absorbing ions come from a very narrow temperature range around their peak temperature $T_{peak}$. 

\placetable{t5}

However, we also find that there exists a non-Gaussian tail in each plot. This tail can be well fitted by a lognormal distribution (the solid line)
\begin{equation}
f \propto \exp \left[-\log^{2}(b/\bar{b}_{ln})/2\sigma_{ln}^{2}\right]
\end{equation} Here $\bar{b}_{ln}$ and $\sigma_{ln}$ are the mean velocity and the standard deviation of this lognormal distribution. Both fitting parameters are listed in columns 4 and 5 of Table~\ref{t5}. We find this distribution naturally cuts off at low $b$, and also provide an excellent fit to the long tail. This non-Gaussian tail implies that the Hubble flow and peculiar velocity contribute a significant part to those ions with high $b$ values.
 
To associate the Doppler $b$-parameter with column density for each line, we show the scatter plots of $b$ vs. $N^{i}$ for all six ions (Figure~\ref{f18}). Each plot displays a left border and a lower cutoff. The left border is an artifact of selecting a threshold $F_{t}$ in synthesizing spectrum. For instance, in the case of \ion{O}{7}, given $F_{t} = 10^{-5}$, equation [\ref{eq:Nb}] imposes (the dotted line in Figure~\ref{f18})
\begin{equation}
b < 2250 \left(\frac{N^{i}}{10^{12}}\right)\ km\ s^{-1}
\end{equation}

\placefigure{f18}

To understand the lower cutoff $b_{min}$, we need to understand where the lines with the lower cutoff come from. From Figure~(\ref{f18}) we know that for lines with a given column density $N^{i}$, they have different Doppler-$b$ values. These lines come from either virialized halos (the intersection regions) or filaments. Since filaments are shock-heated (see, e.g., \citealp{dco00}) and have a much larger $b_{prof}$, we assume that the lines with the lowest Doppler-$b$ values ($b_{min}$) come from virialized halos, and for these lines $b_{min} \approx b_{T} \propto T^{1/2}_{vir}$ where $T_{vir}$ is the virial temperature of the halo. However, even these lines could come from different halos with different virial temperatures: they could come from either the outskirts of a larger halo, or the central region of a smaller halo. According to the mass-temperature relationship we know that a smaller halo has a lower virial temperature and so has a smaller $b$ value. We conclude that those lines with lower cutoffs ($b_{min}$) come from the center of virialized halos. In Figure~\ref{f18} we plot $b_{T}$ vs. the central column densities of all virialized halos (the solid lines), based on our analytic model. We find that this gives a quite good fit to the lower cutoff in all six plots. The turnover at higher $b$ is due to the fact that the ionization fractions fall at higher temperatures and so do the central ion column densities. 

\section{Summary and Discussion \label{sec:sd}}

\begin{figure}
\plotone{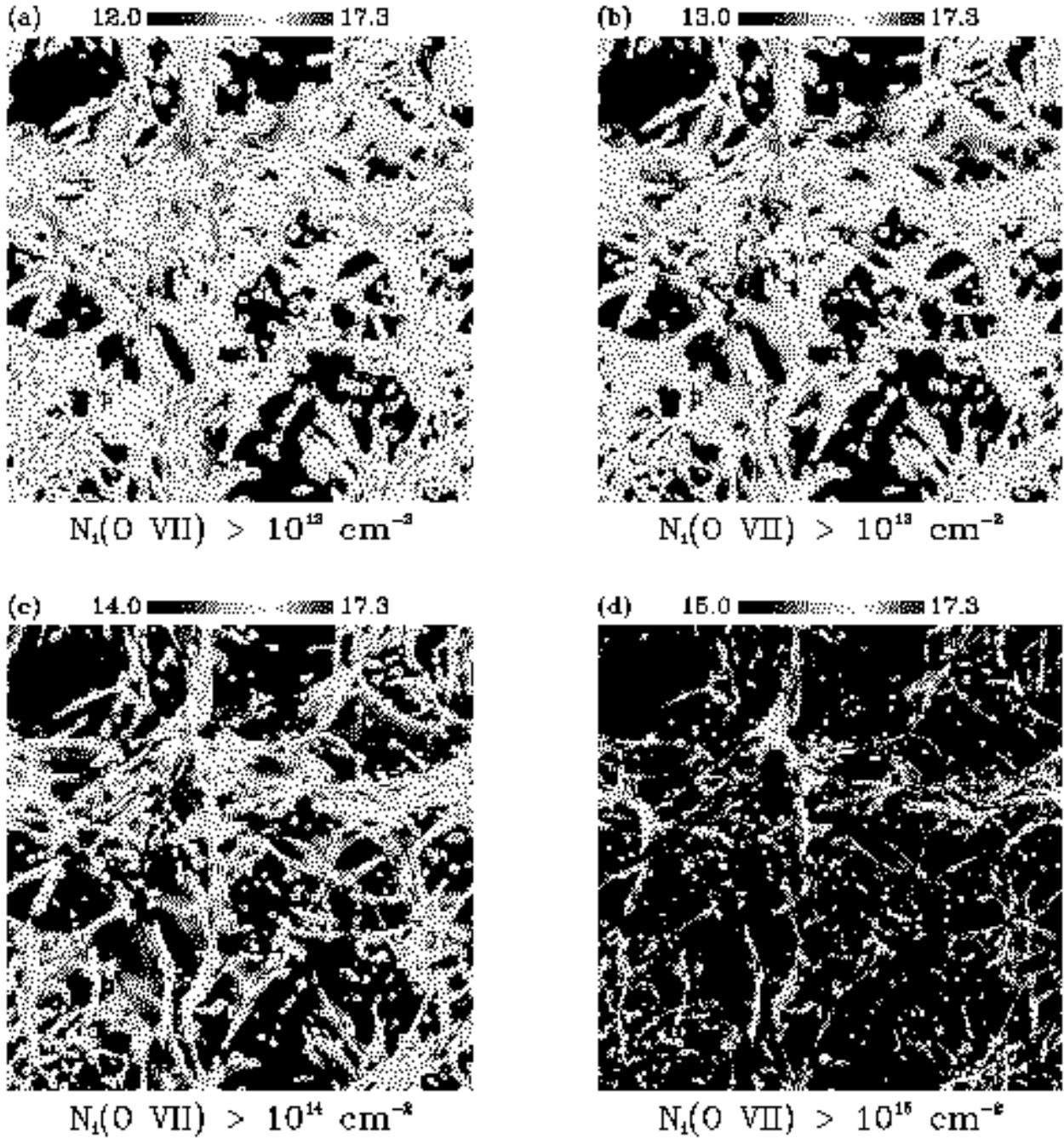}
\caption{The projected density (column density, $N^{i}$) distributions of \ion{O}{7}, plotted in four different panels with $N^{i} > 10^{12}$, $10^{13}$, $10^{14}$ and $10^{15}\ cm^{-2}$, respectively. The color bar on top of each panel indicates the logarithm scale from the minimum to the maximum column densities in each panel. \label{f4}}
\end{figure}

\begin{figure}
\plotone{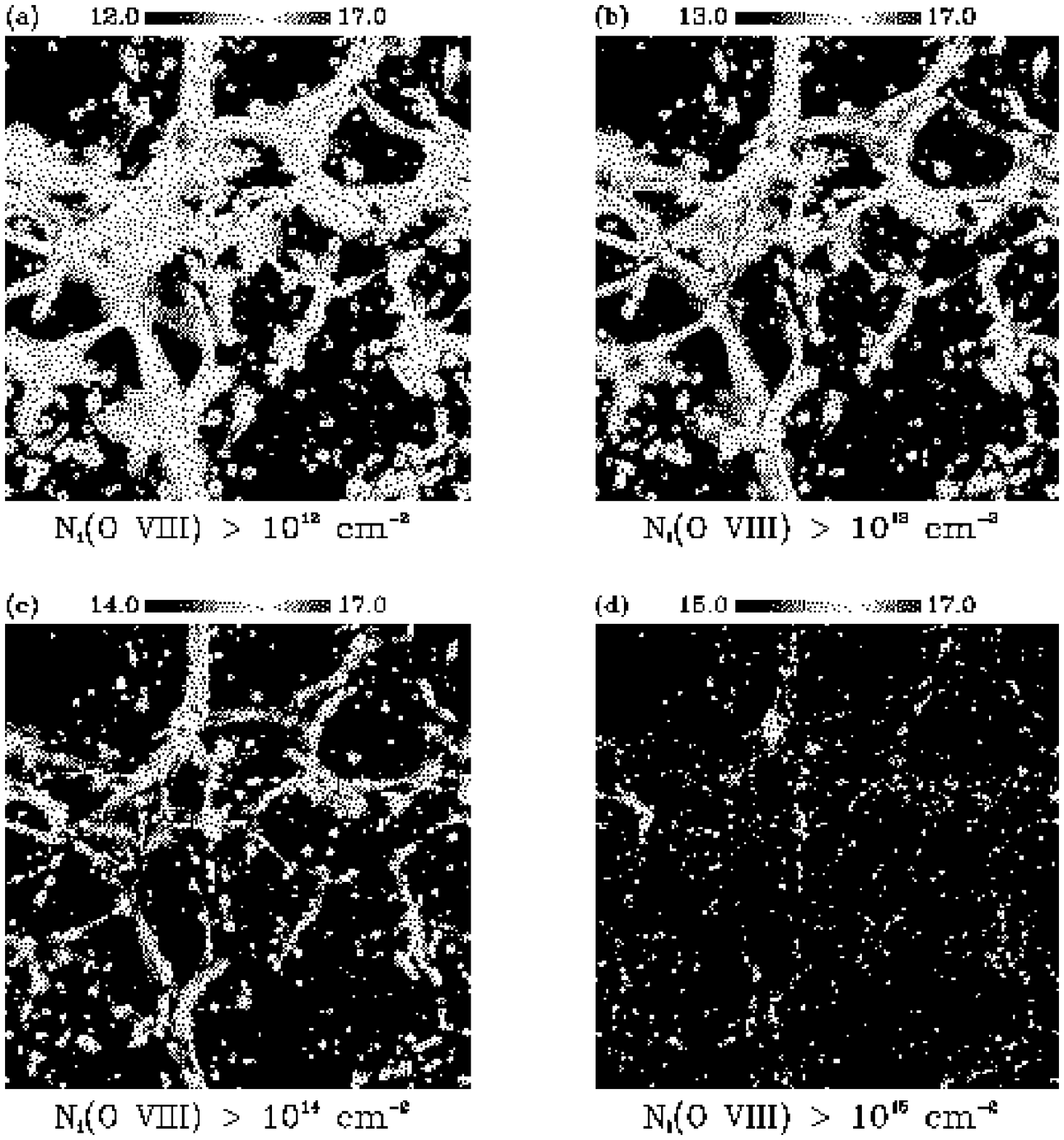}
\caption{The projected column density distributions of \ion{O}{8}. \label{f5}}
\end{figure}

\begin{figure}
\plotone{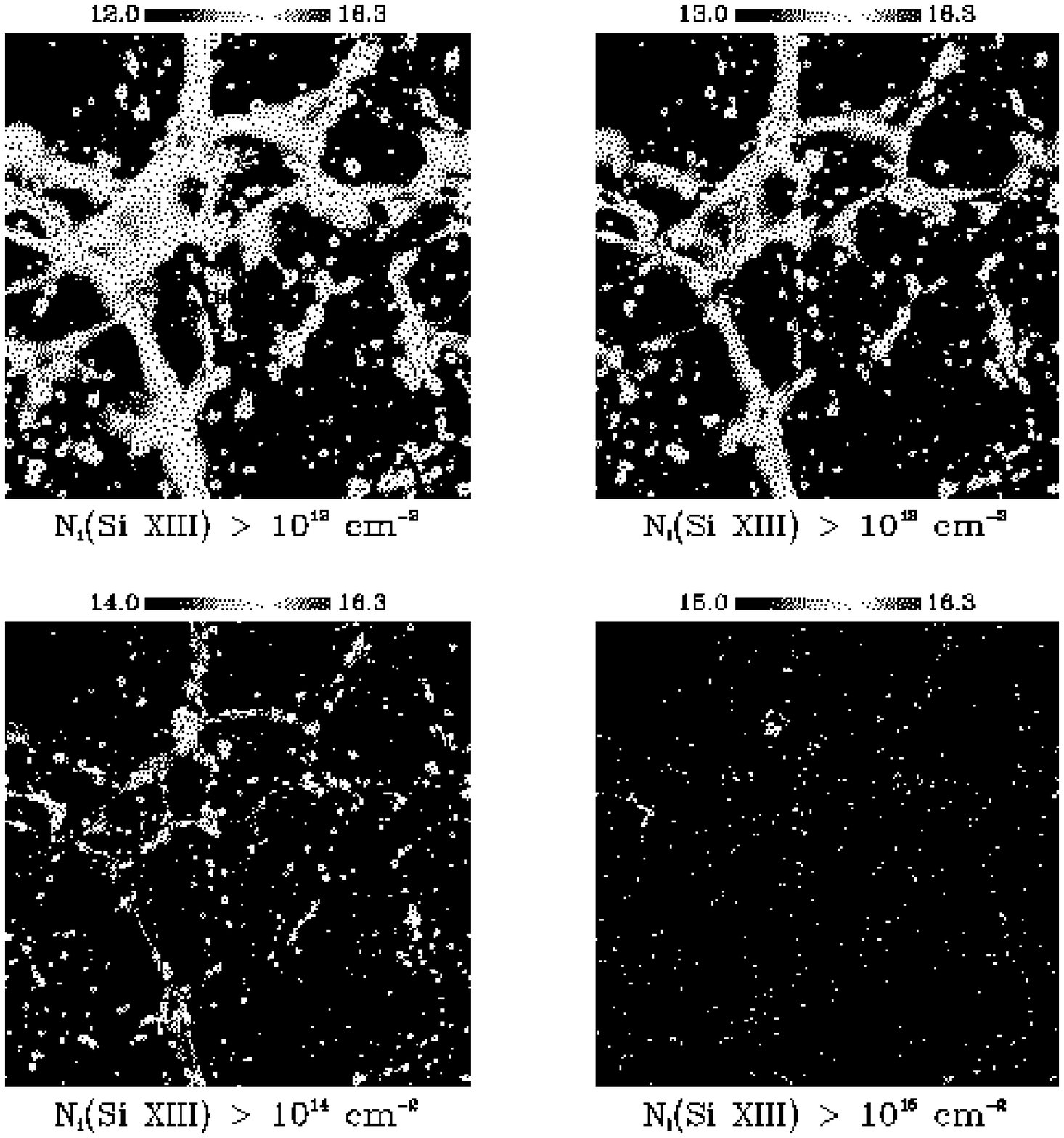}
\caption{The projected column density distributions of \ion{Si}{13}. \label{f6}}\end{figure}

\begin{figure}
\plotone{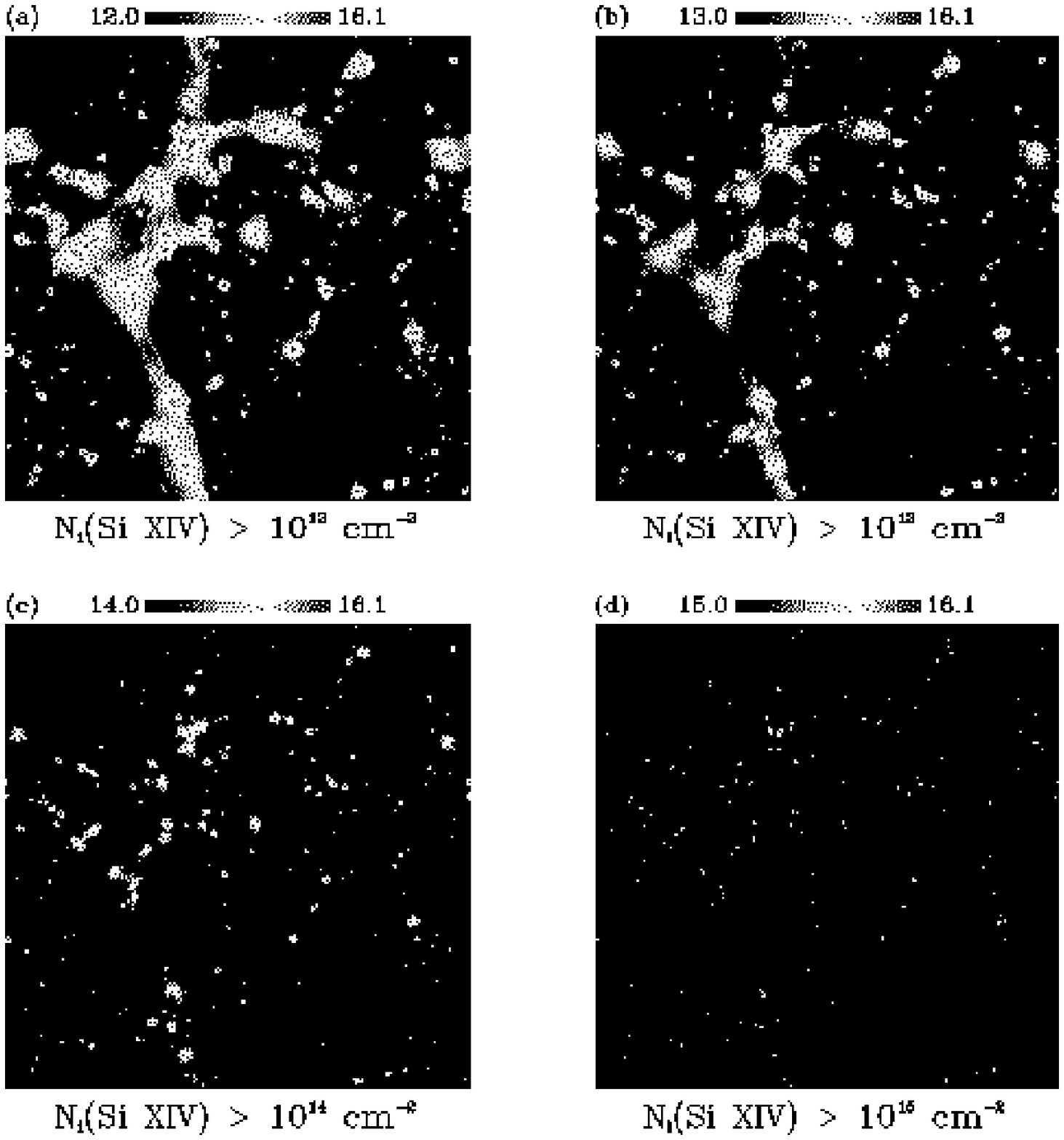}
\caption{The projected column density distributions of \ion{Si}{14}. \label{f7}}
\end{figure}

\begin{figure}
\plotone{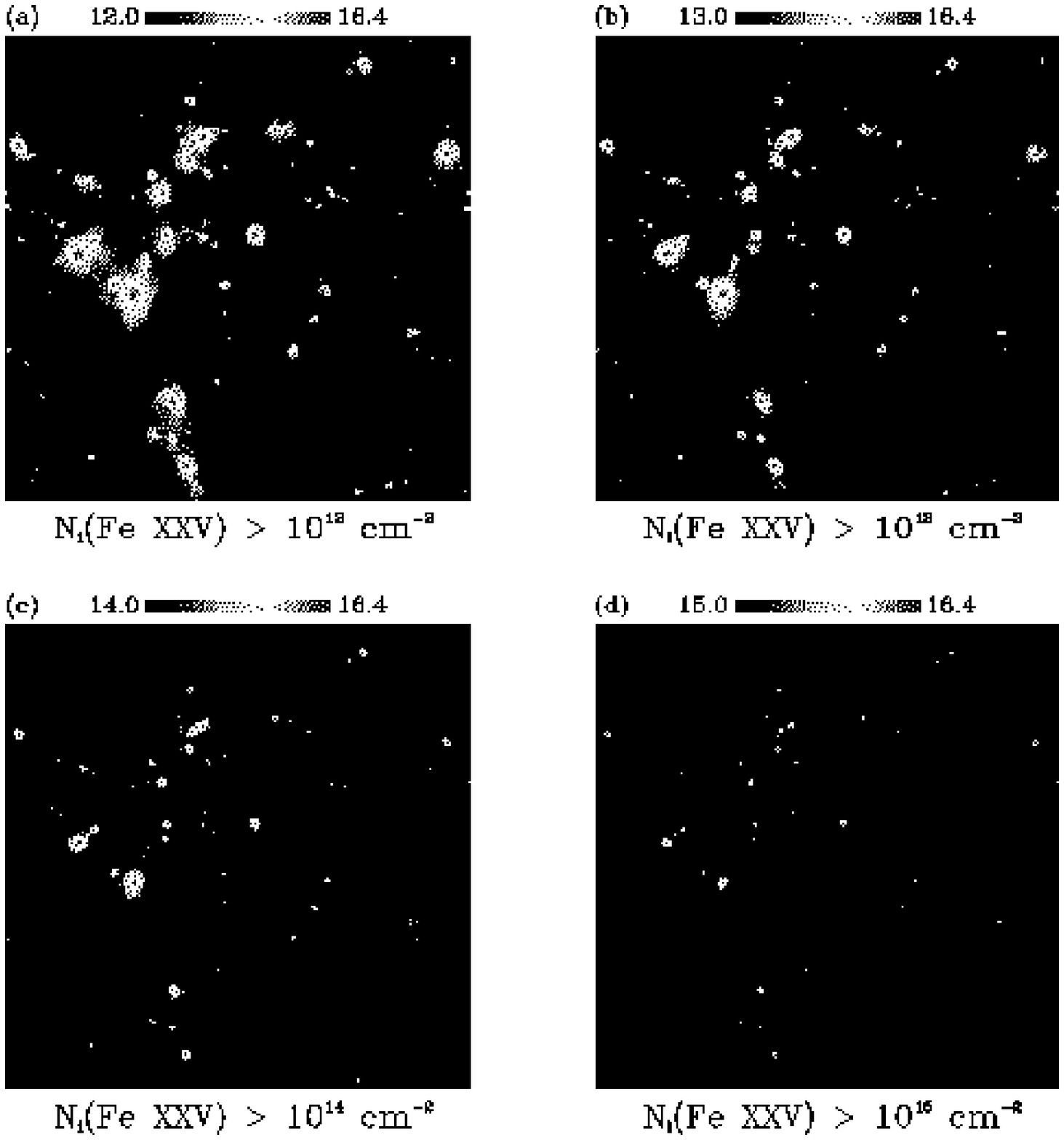}
\caption{The projected column density distributions of \ion{Fe}{25}. \label{f8}}
\end{figure}

\begin{figure}
\plotone{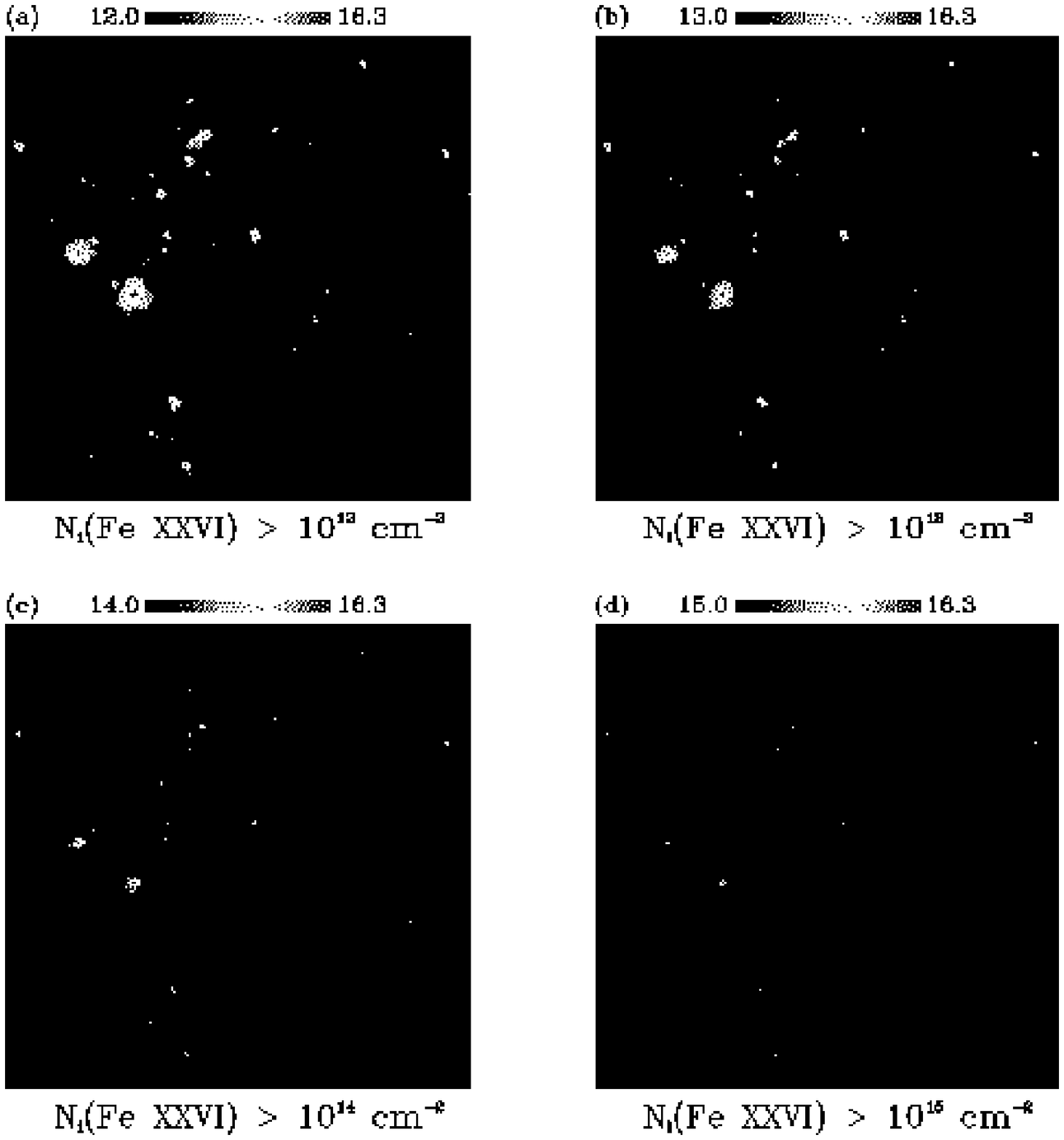}
\caption{The projected column density distributions of \ion{Fe}{26}. \label{f9}}
\end{figure}

\begin{figure}
\plotone{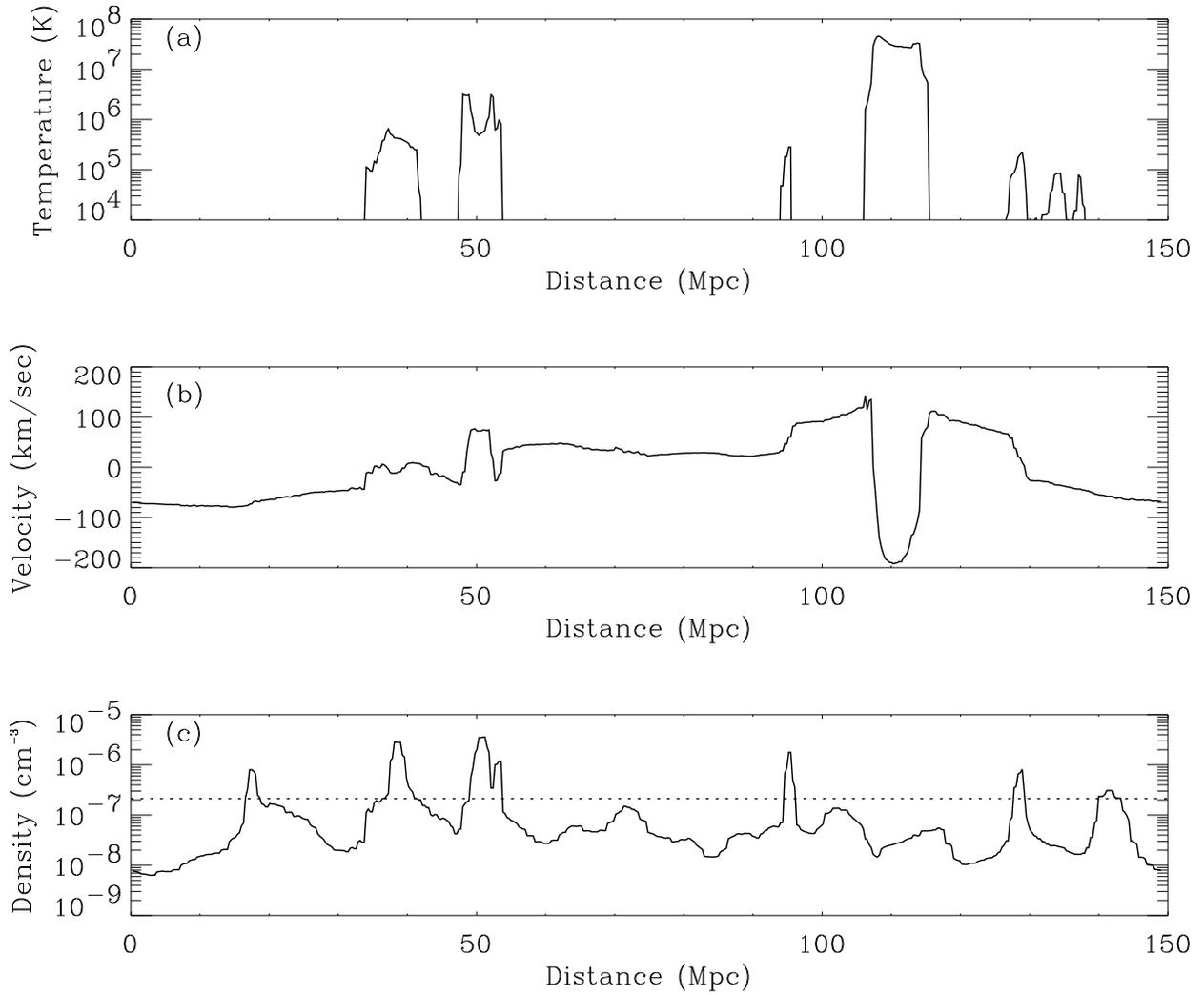}
\caption{(a) Temperature, (b) peculiar velocity (coherent or turbulent) and (c) gas density distributions along one LOS. The dotted line in (c) is the mean baryon density of the universe. \label{f10}}
\end{figure}

\begin{figure}
\plotone{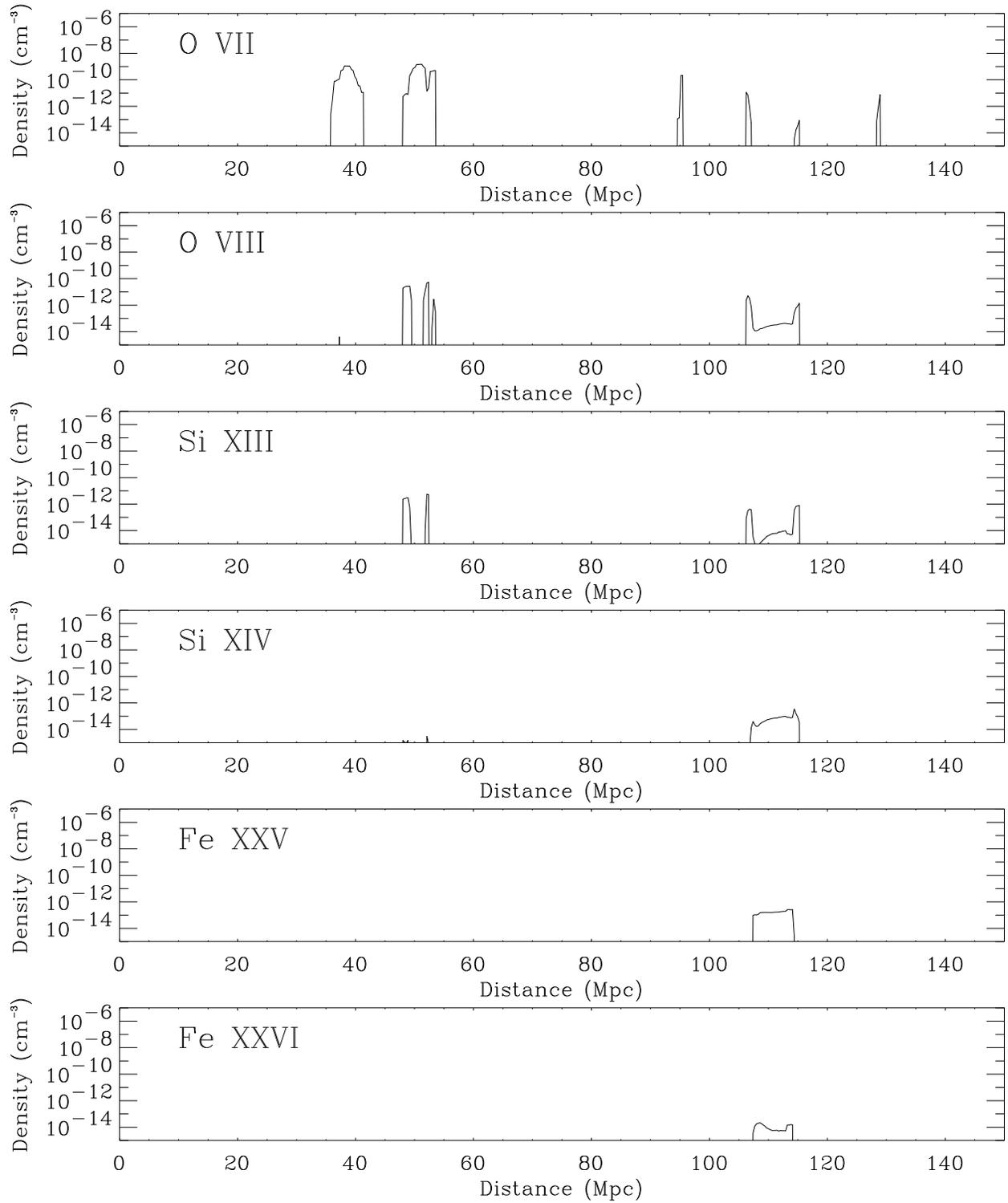}
\caption{The number density distributions along one LOS for six ions. \label{f11}}
\end{figure}

\begin{figure}
\plotone{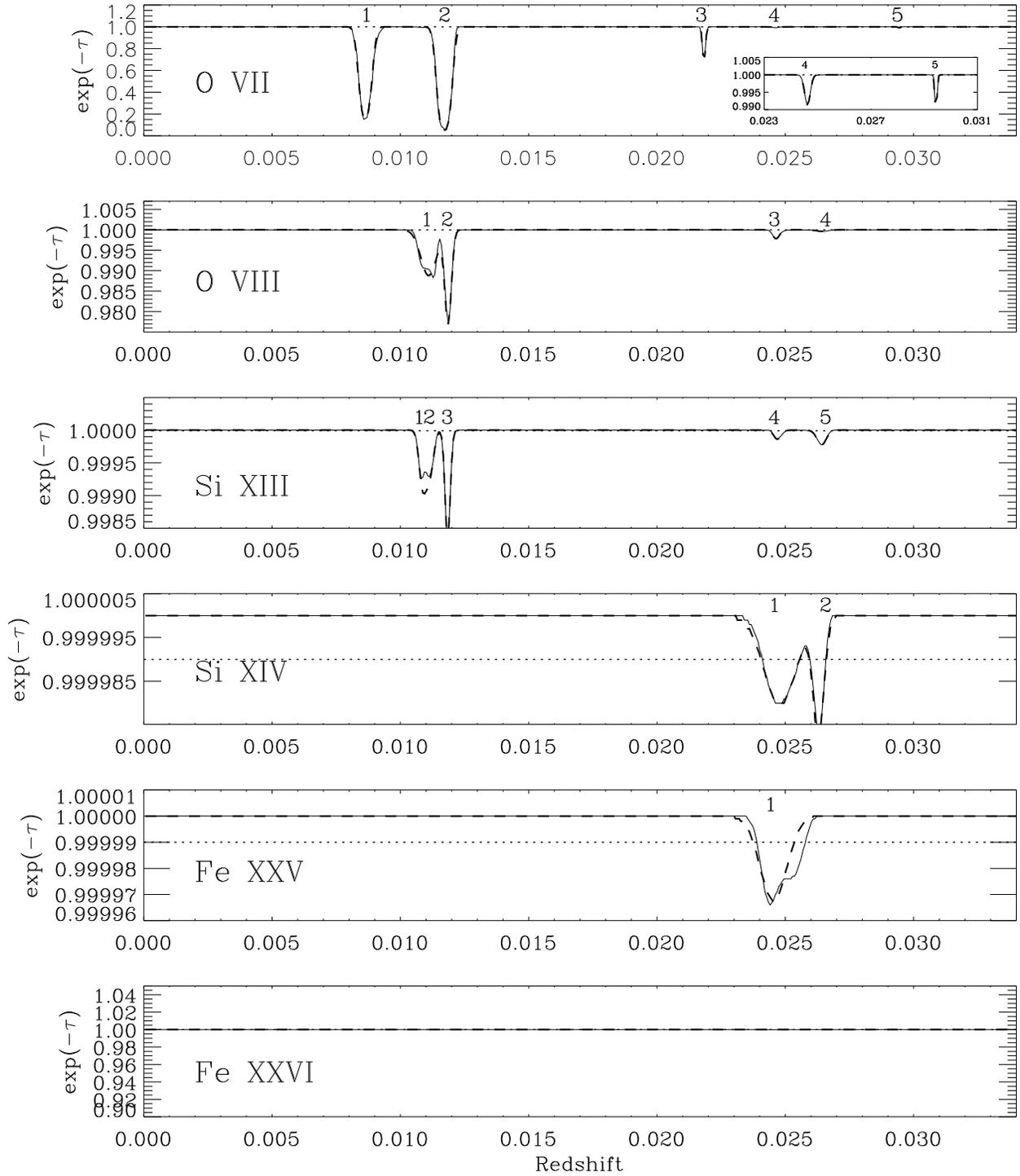}
\caption{Synthesized absorption spectra. The solid lines are the spectra from the numerical simulation and the dashed lines are from the fitted data by the ``{\it threshold-based}'' method. Here $F_{t} = 10^{-5}$. Dotted lines indicate $1-F_{t}$. \label{f12}}
\end{figure}

\begin{figure}
\plotone{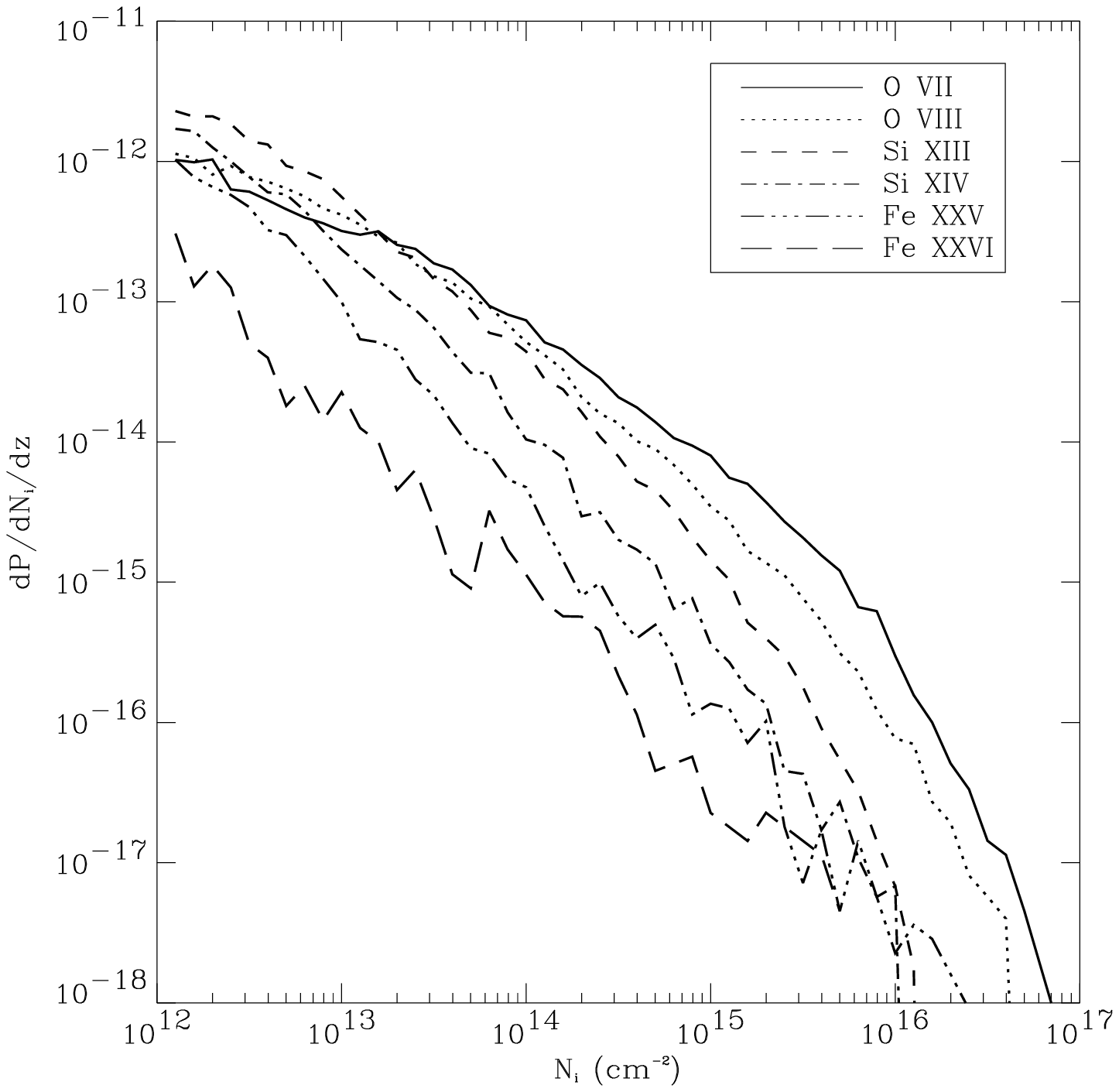}
\caption{The distribution functions from \ion{O}{7} to \ion{Fe}{26} for $F_{t} = 10^{-5}$. \label{f13}}
\end{figure}

\begin{figure}
\plotone{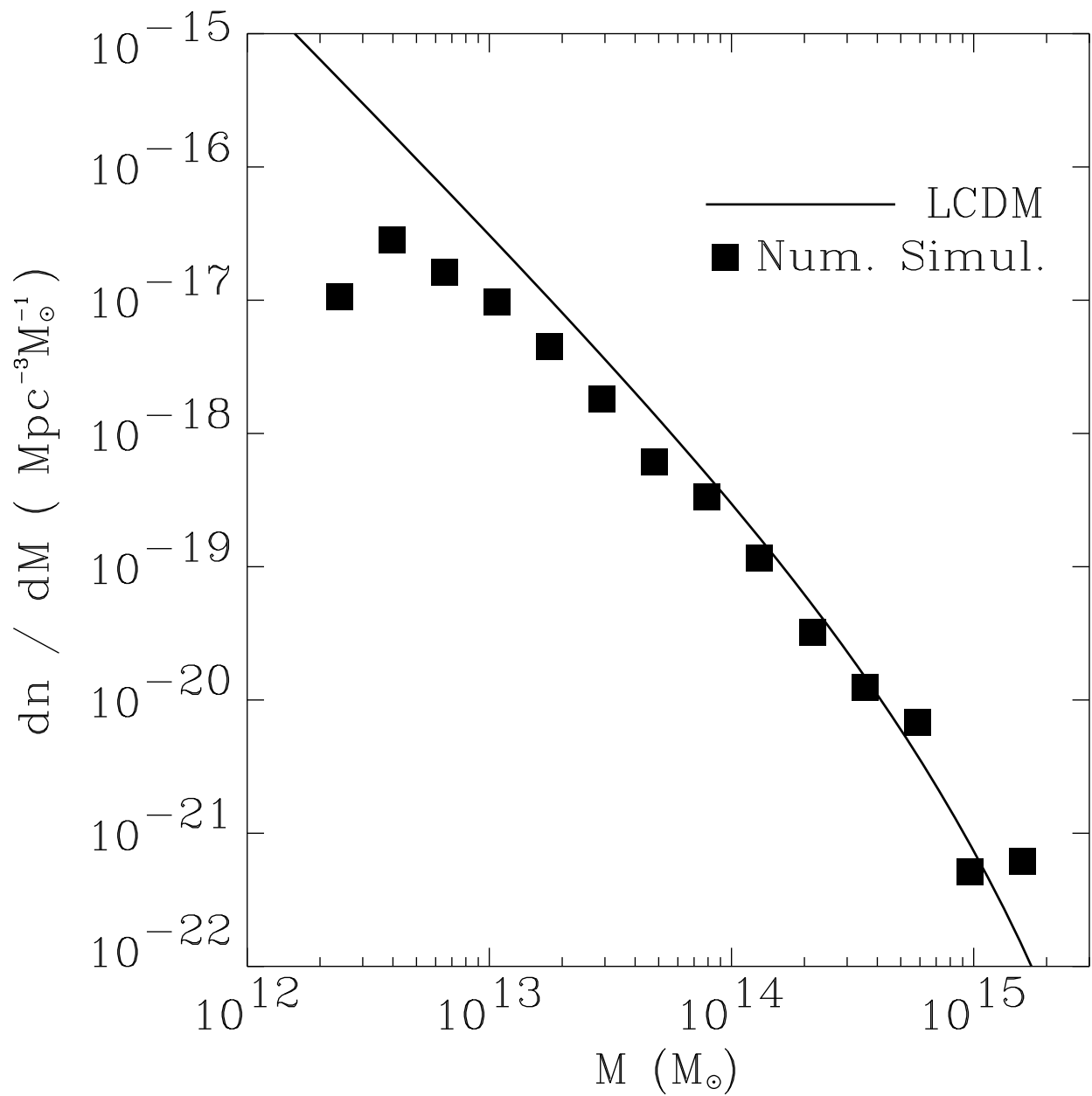}
\caption{The comparison between the numerical simulation (the filled square) and the analytic model (the sold line) for the number density of virialized objects. The analytic calculation is based on the LCDM cosmological model we adopt in the numerical simulation. \label{f14}}
\end{figure}

\begin{figure}
\plotone{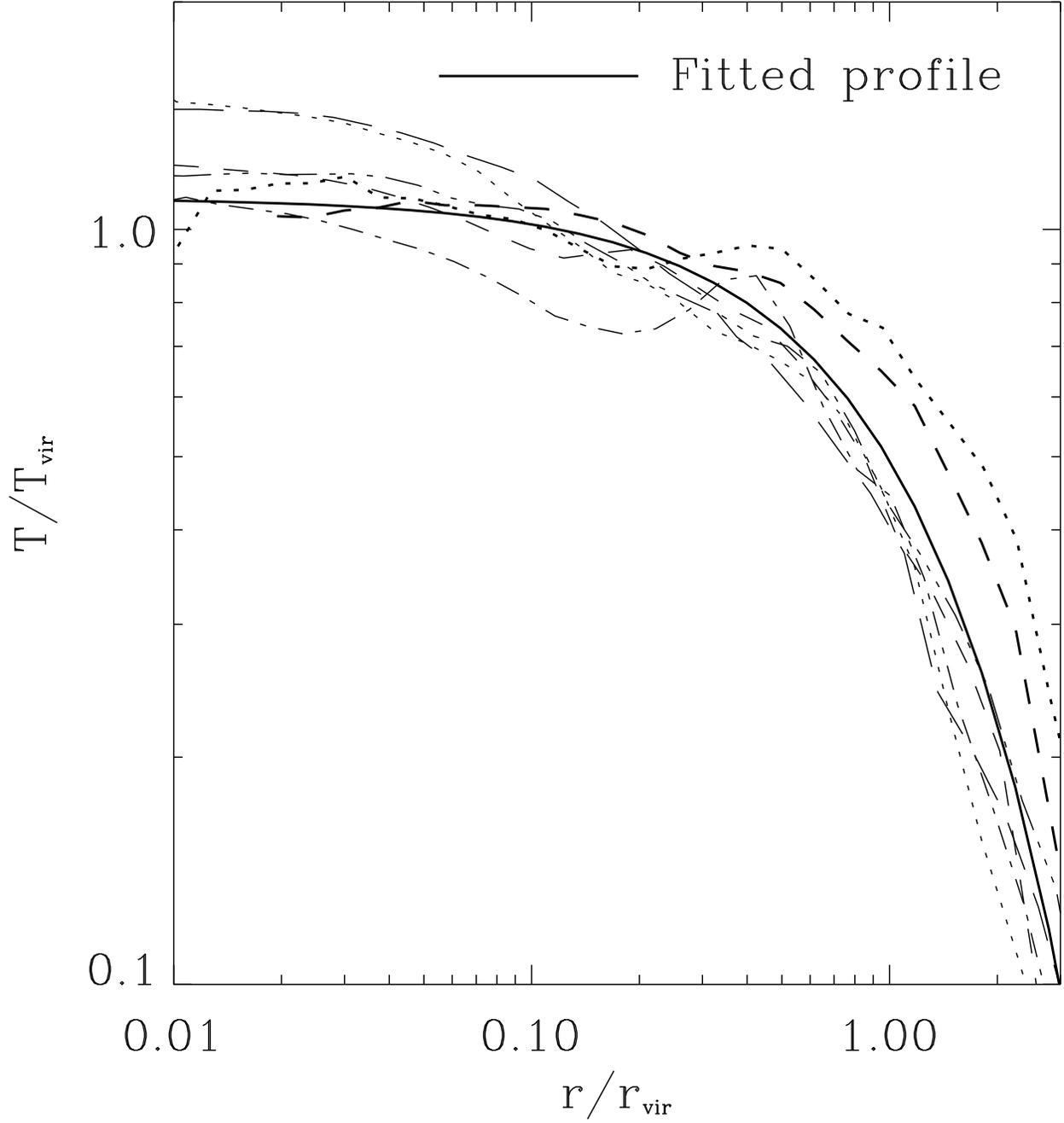}
\caption{The normalized radial temperature profile for several identified clusters in the simulation. The solid line is a fit to simulated profiles, $T_{vir}$ and $r_{vir}$ are the virial temperature and radius of clusters. \label{f15}}
\end{figure}

\begin{figure}
\plotone{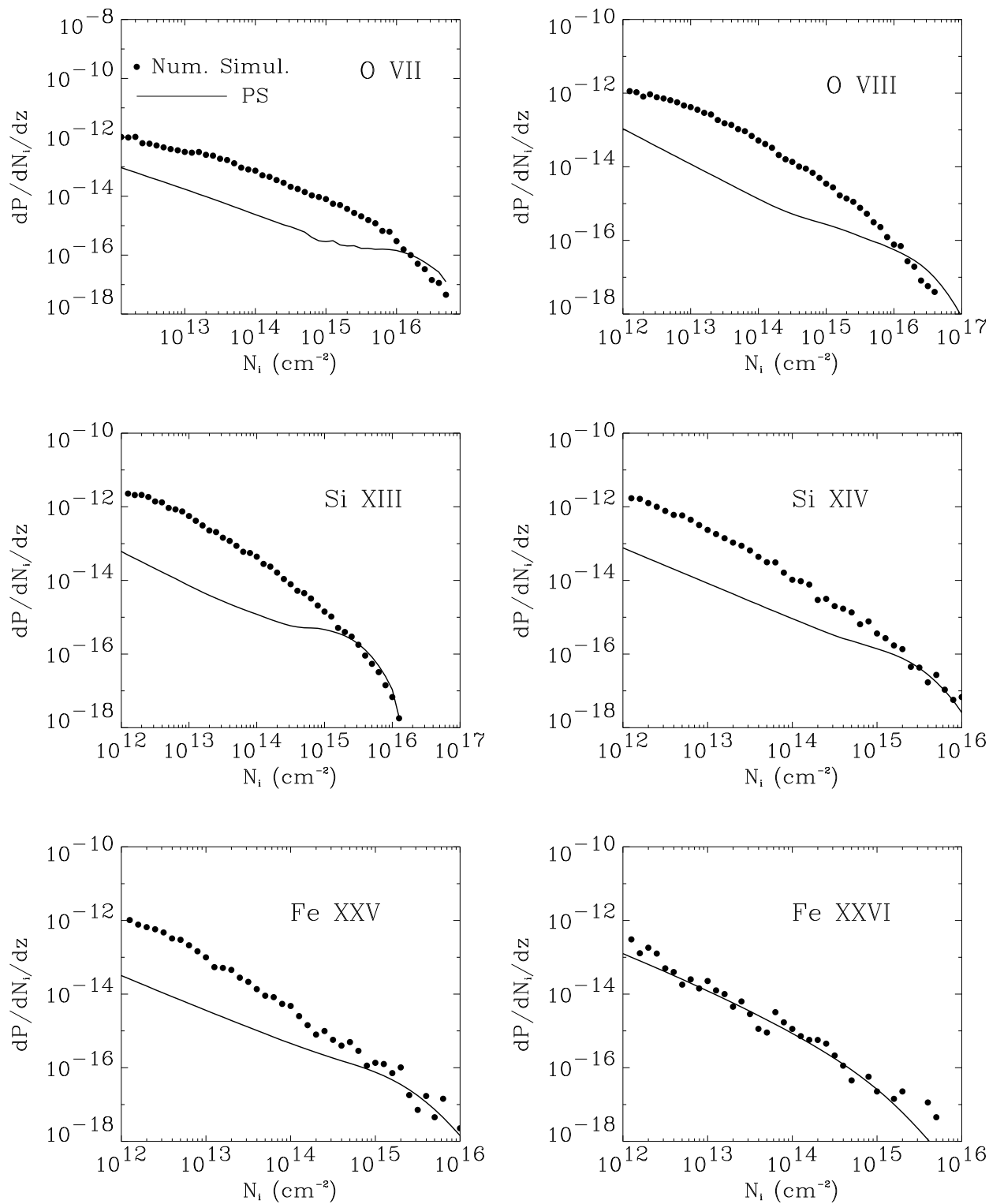}
\caption{The XFDF predicted from the Press-Schechter approach (solid lines) and numerical simulation (filled-dotted lines). \label{f16}}
\end{figure}

\begin{figure}
\plotone{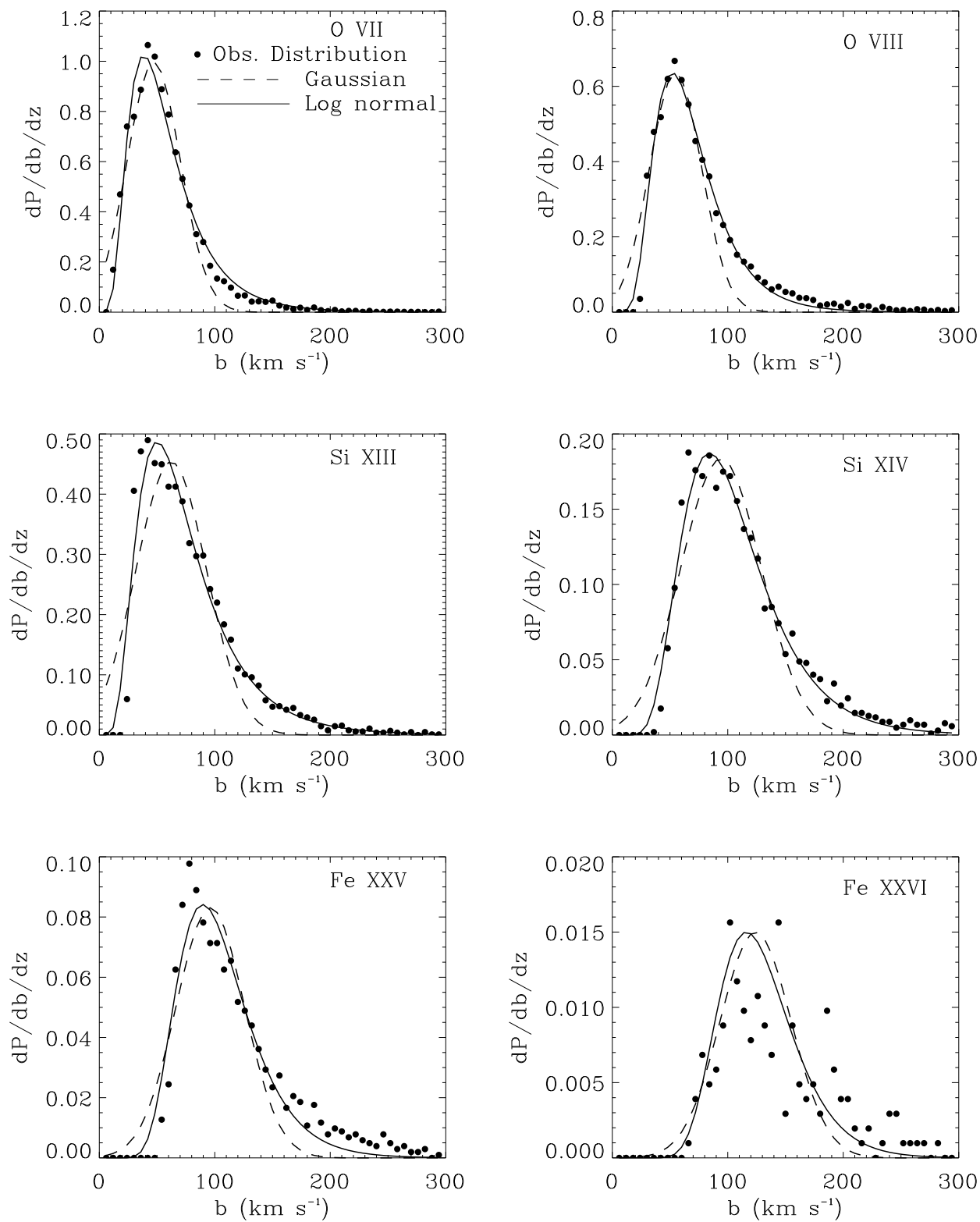}
\caption{The $b$-parameter distribution function for six ions (the dotted lines), fitted with Gaussian distributions (the dashed lines) and log normal distributions (the solid lines). See Table~\ref{t5} for fitting parameters. \label{f17}}
\end{figure}

\begin{figure}
\plotone{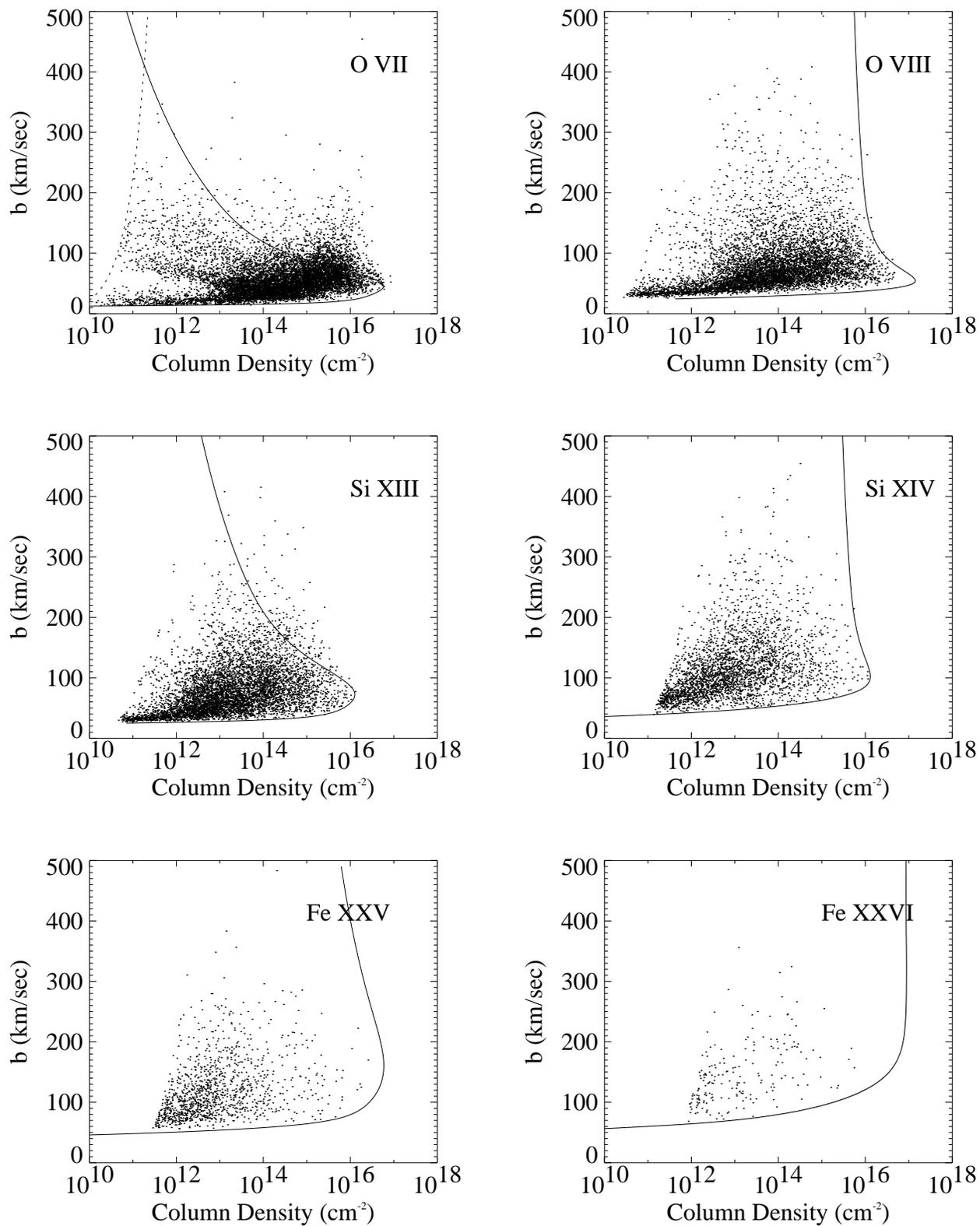}
\caption{The scatter plots of $b$ vs. $N^{i}$ for all six ions. Each plot displays a left border, which gives a maximum $b$ value ($b_{max}$), and a lower cutoff, which gives a minimum $b$ value ($b_{min}$). \label{f18}}
\end{figure}

In this paper we have performed a hydrodynamic simulation based on a LCDM model. By drawing random LOS and synthesizing the X-ray spectrum for each LOS, we obtained the XFDFs for hydrogen and helium-like O, Si, Fe. We have investigated the column density distributions of highly-ionized metals by comparing the numerical simulation to the analytic method described in \citet{fca00}. We also discuss the distribution of Doppler $b$-parameters. Our main conclusions can be summarized as follows:

\begin{enumerate}

\item Ions with low peak temperatures such as \ion{O}{7} and \ion{O}{8} trace small halos (galaxy groups) and filaments, while high peak ions such as \ion{Fe}{25} and \ion{Fe}{26} are only observed in the largest objects (galaxy clusters). For all ions, the strongest signal come from virialized halos; however, filaments show up as lower column density absorbers (see Figure~\ref{f4} to \ref{f9}).

\item The low peak temperature ions are more abundant than ions with high peak temperatures. This can be understood from numerical simulations that most of the baryons have temperatures ranging from $10^{5}-10^{7}$ K and the baryon fraction decreases as the temperature increases. This implies that from the point view of observation, ions with low peak temperatures such as \ion{O}{7} and \ion{O}{8} should be more easily observed.

\item We compared the column density distribution of absorbers predicted by the simulations against an analytic model developed in \citet{fca00} and found that the numerical simulation confirms our predictions: ions with high peak temperatures ($T > 10^{8}$K) follow the distributions of collapsed halos, which can be described by the Press-Schechter distribution; for ions with low peak temperatures, those lines with high column density come from virialized structures, while those with low column density are distributed in diffuse filaments which are not well described by the PS distribution.

\item Higher excitation ions produce lines which are generally wider (as determined by the Doppler $b$-parameter) than low excitation lines, although there is a wide distribution in all cases. This distribution is well fit by a log normal distribution. The sharp lower cutoff of this distribution is provided by the fact that there is a minimum temperature required to create the ion in the first place. Peculiar velocities in the gas also contribute significantly, providing a non-Gaussian tail at the high-$b$ end.

\end{enumerate}

The largest uncertainty in spectral synthesis comes from the metallicity: we did not include any gas processes in our numerical simulation. Although these gas processes are not important in heating and cooling gas with $T>10^{5}$ K, the IGM metals enriched by these processes are crucial in UV/X-ray observations: it is these metals which produces detectable features. We assume a uniform metallicity in generating the synthesized spectrum, however, we would expect that metallicity should be determined by the metal enrichment history of the IGM, which in turn depends on a variety of sources: the UV/X-ray background radiation, supernova input, etc, and these are likely to be different in low and high density regions. It is important to incorporate all these processes to obtain a self-consistent result of metal distributions and ionization structures in the IGM. This is the goal of our future numerical simulations.

\acknowledgments{TF thanks the support from MIT/CXC team. This work is supported in part by contracts NAS 8-38249 and SAO SV1-61010. Support for GLB was provided by NASA through Hubble Fellowship grant HF-01104.01-98A from the Space Telescope Science Institute, which is operated by the Association of Universities for Research in Astronomy, Inc., under NASA contract NAS 6-26555.}

\clearpage

\begin{deluxetable}{lccccc}
\tablewidth{0pt}
\tablecaption{Ion Parameters \label{t1}}
\tablehead{
\colhead{Ion} &  \colhead{Energy} & \colhead{Wavelength} &
\colhead{Oscillation} & \colhead{Metallicity\tablenotemark{a}} & 
\colhead{Solar} \\ \colhead{} &  \colhead{($keV$)} & \colhead{($\AA$)} & 
\colhead{Strength} & \colhead{} & \colhead{Abundance\tablenotemark{b}}
}
\startdata
\ion{O}{7}   & 0.57 & 21.6019 & 0.696 & 0.5 & $8.53 \times 10^{-4}$\nl
\ion{O}{8}   & 0.65 & 18.9602 & 0.416 & 0.5 & $8.53 \times 10^{-4}$\nl
\ion{Si}{13} & 1.87 & 6.6480  & 0.757 & 0.5 & $3.58 \times 10^{-5}$\nl
\ion{Si}{14} & 2.01 & 6.1822  & 0.416 & 0.5 & $3.58 \times 10^{-5}$\nl
\ion{Fe}{25} & 6.70 & 1.8504  & 0.798 & 0.3 & $3.23 \times 10^{-5}$\nl
\ion{Fe}{26} & 6.97 & 1.7798  & 0.416 & 0.3 & $3.23 \times 10^{-5}$\nl
\enddata
\tablenotetext{a}{Relative to solar abundance $Z_{\odot}$.}
\tablenotetext{b}{Relative to hydrogen number density.}
\end{deluxetable}
\clearpage

\begin{deluxetable}{lcc}
\tablewidth{0pt}
\tablecaption{Minimum Column Densities for \ion{O}{7} and \ion{O}{8} \label{t2}}
\tablehead{
\colhead{Threshold ($F_{T}$)} & \colhead{\ion{O}{7} ($cm^{-2}$)}& \colhead{\ion{O}{8} ($cm^{-2}$)}}
\startdata
$10^{-5}$ & $1.43 \times 10^{10}$ & $2.75 \times 10^{10}$ \nl
$10^{-1}$ & $1.34 \times 10^{14}$ & $4.28 \times 10^{14}$ \nl
0.2 & $2.35 \times 10^{14}$ & $8.37 \times 10^{14}$ \nl
0.5 & $6.82 \times 10^{14}$ & $1.96 \times 10^{15}$ \nl
0.8 & $1.13 \times 10^{15}$ & $3.75 \times 10^{15}$ \nl
\enddata
\end{deluxetable}
\clearpage

\begin{deluxetable}{cccccc}
\tablewidth{0pt}
\tablecaption{Resonant Absorption Lines \tablenotemark{a}\label{t3}}
\tablehead{
\colhead{Ion} & {Line} & \colhead{z} & \colhead{N ($cm^{-2}$)} &
\colhead{$EW\ (eV)$} & \colhead{$b\ (km/s)$} 
}
\startdata
\ion{O}{7} & 1 & $8.57\times10^{-3}$ & $6.40\times10^{15}$ & $2.72\times10^{-1}$ & 76.05 \\
     	   & 2 & $1.17\times10^{-2}$ & $1.04\times10^{16}$ & $3.46\times10^{-1}$ & 79.18 \\
           & 3 & $2.19\times10^{-2}$ & $3.84\times10^{14}$ & $2.42\times10^{-2}$ & 23.00 \\
           & 4 & $2.46\times10^{-2}$ & $1.74\times10^{13}$ & $1.23\times10^{-3}$ & 44.93 \\
           & 5 & $2.94\times10^{-2}$ & $7.76\times10^{12}$ & $5.41\times10^{-4}$ & 16.88 \\
\cline{1-6}
\ion{O}{8} & 1 & $1.13\times10^{-2}$ & $1.23\times10^{14}$ & $5.42\times10^{-3}$ & 128.90 \\
           & 2 & $1.19\times10^{-2}$ & $1.05\times10^{14}$ & $4.57\times10^{-3}$ &  53.98 \\
           & 3 & $2.46\times10^{-2}$ & $1.03\times10^{13}$ & $4.37\times10^{-4}$ &  55.61 \\
           & 4 & $2.64\times10^{-2}$ & $2.83\times10^{12}$ & $1.20\times10^{-4}$ &  89.50 \\
\cline{1-6}
\ion{Si}{13} & 1 & $1.08\times10^{-2}$ & $5.75\times10^{12}$ & $4.63\times10^{-4}$ & 56.14 \\
             & 2 & $1.11\times10^{-2}$ & $6.36\times10^{12}$ & $5.11\times10^{-4}$ & 65.85 \\
             & 3 & $1.18\times10^{-2}$ & $9.45\times10^{12}$ & $7.57\times10^{-4}$ & 42.65 \\
             & 4 & $2.46\times10^{-2}$ & $1.13\times10^{12}$ & $8.73\times10^{-5}$ & 60.77 \\
             & 5 & $2.64\times10^{-2}$ & $2.27\times10^{12}$ & $1.73\times10^{-4}$ & 75.55 \\
\cline{1-6}
\ion{Si}{14} & 1 & $2.47\times10^{-2}$ & $1.41\times10^{12}$ & $5.98\times10^{-5}$ & 267.95 \\
             & 2 & $2.62\times10^{-2}$ & $6.22\times10^{11}$ & $2.63\times10^{-5}$ & 90.47 \\
\cline{1-6}
\ion{Fe}{25} & 1 & $2.44\times10^{-2}$ & $3.24\times10^{12}$ & $1.93\times10^{-6}$ & 220.38 \\
\enddata
\tablenotetext{a}{From a sample LOS, see Figure~\ref{f12}.}
\end{deluxetable}
\clearpage

\begin{deluxetable}{lcccccc}
\tablecaption{Line Properties \label{t4}}
\tablehead{
\colhead{Ion} & \colhead{Line Number} & \colhead{$N_{min}$} &
 \colhead{$N_{max}$} & \colhead{$b_{min}$} & \colhead{$b_{max}$}\\
\colhead{} & \colhead{} & \colhead{($cm^{-2}$)} & \colhead{($cm^{-2}$)} & \colhead{($km/s$)} & \colhead{($km/s$)}
} 
\startdata
\ion{O}{7} & 10222 & $1.43\times10^{10}$ & $8.39\times10^{16}$ & 13.43 & 906.70 \\
\ion{O}{8} & 7078  & $2.75\times10^{10}$ & $4.57\times10^{16}$ & 25.73 & 829.26 \\
\ion{Si}{13} & 6246 & $4.67\times10^{10}$ & $1.30\times10^{16}$ & 25.81 & 697.71 \\
\ion{Si}{14} &  2960 & $1.34\times10^{11}$ & $1.13\times10^{16}$ & 39.30 & 454.34 \\
\ion{Fe}{25} & 1192& $2.86\times10^{11}$ & $2.56\times10^{16}$ & 56.36 & 483.18 \\
\ion{Fe}{26} & 194 & $8.51\times10^{11}$ &  $5.51\times10^{15}$ & 68.63 & 356.05 \\ 
\enddata
\end{deluxetable}
\clearpage

\begin{deluxetable}{lcccc}
\tablewidth{0pt}
\tablecaption{Doppler Parameter Distributions \label{t5}}
\tablehead{
\colhead{} & \multicolumn{2}{c}{Gaussian} & \multicolumn{2}{c}{Log normal} \\
\cline{2-3} \cline{4-5} \\
\colhead{Ion} & \colhead{$\bar{b}$} & \colhead{$\sigma$} &
\colhead{$\bar{b}_{ln}$} & \colhead{$\sigma_{ln}$}
} 
\startdata
\ion{O}{7} & 47.17 & 23.06 & 38.52 & 0.23 \\
\ion{O}{8} & 54.84 & 22.47 & 51.97 & 0.19 \\
\ion{Si}{13} & 62.29 & 30.48 & 50.03 & 0.22 \\
\ion{Si}{14} & 94.18 & 34.85 & 84.02 & 0.17 \\
\ion{Fe}{25} & 96.14 & 29.63 & 89.51 & 0.14 \\
\ion{Fe}{26} & 124.10 & 30.76 & 116.17 & 0.11 \\
\enddata
\end{deluxetable}
\clearpage

\clearpage

\clearpage

\clearpage

\clearpage

\clearpage

\clearpage

\clearpage

\clearpage

\clearpage

\clearpage

\clearpage

\clearpage

\clearpage

\clearpage

\clearpage

\clearpage

\clearpage

\clearpage

\end{document}